\documentclass{aa}

\usepackage{graphicx}
\usepackage[varg]{txfonts}

\usepackage{natbib}
\usepackage{hyperref}
\bibpunct{(}{)}{;}{a}{}{,} % to follow the A&A style

\begin{document}

\title{The role of convection, overshoot, and gravity waves for the transport
  of dust in M dwarf and brown dwarf atmospheres}

\author{
Bernd Freytag \inst{1},
France Allard \inst{1,2},
Hans-G{\"u}nter Ludwig \inst{3},
Derek Homeier \inst{4},
Matthias Steffen \inst{5}
       }

\authorrunning{Freytag et al.}
\titlerunning{Convection, overshoot, and gravity waves in low-mass dwarfs}

\offprints{B. Freytag}

\institute{
Centre de Recherche Astrophysique de Lyon,
UMR 5574: CNRS, Universit\'e de Lyon,
\'Ecole Normale Sup\'erieure de Lyon,
46 all\'ee d'Italie, F-69364 Lyon Cedex 07, France
  \and
Institut d'Astrophysique de Paris, UMR 7095: CNRS, Universit\'e Pierre et Marie Curie-Paris 6, 98bis boulevard Arago, 75014 Paris, France  
  \and
Observatoire de Paris-Meudon,
GEPI-CIFIST, F-92195 Meudon, France
  \and
Institut f{\"u}r Astrophysik G{\"o}ttingen,
Georg-August-Universit{\"a}t,
Friedrich-Hund-Platz 1,
D-37077 G{\"o}ttingen, Germany
  \and
Astrophysikalisches Institut Potsdam,
An der Sternwarte 16,
D-14482 Potsdam, Germany
          }

\date{Received ...; accepted ...}

%###############################################################################
% \abstract{}{}{}{}{}
% 5 {} token are mandatory

\abstract
% --- context heading (optional), {} leave it empty if necessary ---
{
Observationally, spectra of brown dwarfs indicate the presence of dust in
their atmospheres while theoretically it is not clear what prevents the dust
from settling and disappearing from the regions of spectrum formation.
Consequently, standard models have to rely on ad hoc assumptions
about the mechanism that keeps dust grains aloft in the atmosphere.
}
% --- aims heading (mandatory) ---
{
We apply hydrodynamical simulations to develop
an improved physical understanding of the mixing properties
of macroscopic flows in M dwarf and brown dwarf atmospheres,
in particular of the influence of the underlying convection zone.
}
% --- methods heading (mandatory) ---
{
We performed two-dimensional radiation hydrodynamics simulations
including a description of dust grain formation and transport
with the CO5BOLD code.
The simulations cover the very top of the convection zone and the photosphere
including the dust layers
for a sequence of effective temperatures between 900\,K and 2800\,K,
all with $\log\,g$=5 assuming solar chemical composition.
}
% --- results heading (mandatory) ---
{
Convective overshoot occurs in the form of
exponentially declining velocities with small scale heights,
so that it affects only the region immediately above the
almost adiabatic convective layers.
From there on, mixing is provided by gravity waves that are
strong enough to maintain thin dust clouds in the hotter models. With
decreasing effective temperature, the amplitudes of the waves become smaller
but the clouds become thicker and develop internal convective flows that are
more efficient in transporting and mixing material than gravity waves.
The presence of clouds often leads to a
highly structured appearance of the stellar surface on short temporal and
small spatial scales (presently inaccessible to observations).
}
% --- conclusions heading (optional), leave it empty if necessary ---
{
We identify convectively excited gravity waves as an essential mixing process
in M dwarf and brown dwarf atmospheres.
Under conditions of strong cloud formation, dust convection is the dominant
self-sustaining mixing component.
}

\maketitle

%#########################################################################################
\section{Introduction}

Brown dwarfs form like stars and evolve as they cool from
stellar-like properties --
M spectral type characterized by molecular hydrogen and water vapor formation,
chromospheric activity, flares, and magnetic spots --
to planet-like properties --
T spectral type characterized by methane absorption,
electron-degenerate core, and maser emission.
With a fully convective interior reaching up to the
atmosphere, and a neutral atmosphere offering little interaction with
magnetic field lines, they retain larger rotational velocities
($\ge 30$ km\,s$^{-1}$, i.e., $P \le 4$ hrs compared to 11 hrs for Jupiter).
This efficiency and
the unusually large extent of the convection zone into the atmosphere (up to
optical depths of $10^{-3}$ for M dwarfs) assigns an important role as a
cooling and contraction evolution regulator to the atmospheres
\citep{Baraffe1995ApJ...446L..35B}. Understanding the atmospheric
properties has therefore implications for
the mass determination of these objects.
Some 700 brown dwarfs have been
found\footnote{Photometry, spectroscopy, and astrometry of M, L, and T dwarfs:
\href{http://spider.ipac.caltech.edu/staff/davy/ARCHIVE/index.shtml}{http://spider.ipac.caltech.edu/staff/davy/ARCHIVE/index.shtml}}
%\citep{DwarfArchives.org2009}
in the solar neighborhood and in star-forming regions since the early 90's
reaching into ever cooler and lower mass regimes
($T_\mathrm{eff} \ge 600$\,K, $M \ge 5\ M_\mathrm{Jup}$).
 
Atmospheric temperatures are sufficiently low
($T_{\rm gas} \le 1800$\,K, $T_{\rm eff} \le 2800$\,K)
for dust particle formation to occur in late-type M dwarfs. 
These grains should sink under the influence of
gravity ($g \approx 10^5$ cm\,s$^{-2}$) into deeper layers and vanish from
the atmosphere, clearing it from condensable material. However,
their near-infrared spectra
can only be reproduced when accounting for
a strong greenhouse effect
(also called a blanketing effect in stellar physics) in the visible layers
\citep{Tsuji1996A&A...305L...1T,
Alexander1997Ap&SS.251..171A,
Ruiz1997ApJ...491L.107R,
Leggett1998ApJ...509..836L,
Leggett2001ApJ...548..908L}.
Classical static model atmospheres have to rely on ad hoc assumptions about the
mechanism that keeps dust from settling, or that brings fresh material
toward the surface allowing new grains to form
\citep{Helling2008MNRAS.tmp.1310H}.
The effects of dust formation on the atmospheres of late-type dwarfs
have been explored by modeling dust formation in chemical equilibrium
with the gas phase using diverse prescriptions of the cloud thickness
\citep{Allard2001ApJ...556..357A,
Tsuji2002ApJ...575..264T,
Burrows2006ApJ...640.1063B,
AckermanMarley2001ApJ...556..872A}.
It has been found that the \texttt{PHOENIX} \emph{Dusty} models
\citep{Allard2001ApJ...556..357A} reproduce the infrared emission of
late-type M to mid-L dwarfs
(i.e.,\ $1700\,\mathrm{K} \le T_{\rm eff} \le 2500$\,K)
\citep{Leggett1998ApJ...509..836L,
Leggett2001ApJ...548..908L,
Ruiz1997ApJ...491L.107R}.
This indicates that dust forms close to equilibrium  in the
infrared-line-forming region of these atmospheres ($\tau \approx 10^{-2}$).  

On the other hand, late-type L and T dwarfs
($T_{\rm eff} \le1400$\,K)
are less affected by photospheric greenhouse effects,
but do show evidence --
in terms of higher CO and lower NH$_3$ and CH$_4$ absorption --
of the dynamical upwelling of N$_2$ and CO gas 
\citep{Saumon2006ApJ...647..552S}.
\citet{Cushing2008ApJ...678.1372C} and
\citet{Stephens2009ApJ...702..154S} fitted a sequence of the
red optical to mid-infrared spectra of early L to mid-T dwarfs using
the model atmospheres of  
\citet{AckermanMarley2001ApJ...556..872A} and
\citet{Saumon2006ApJ...647..552S}.
They demonstrated that cloud opacity, adjusted by a sedimentation
efficiency factor $f_\mathrm{sed}$ in these models, affects the
spectra of all dwarfs up to early T types, and the observed
CO/CH$_4$ and N$_2$/NH$_3$ abundances are indicative of mixing effects
equivalent to eddy diffusion coefficients between 10$^2$ and 10$^6$
cm$^2$\,s$^{-1}$ in all atmospheres. But this analysis was still
not unable to a unique relation between spectral type or
$T_\mathrm{eff}$ and sedimentation efficiency. They also found that
the atmospheric parameters derived from best fits to individual
spectral regions would frequently infer different results,
or be in disagreement with expectations from structural and
evolution models.
Thus none of the classical static models have reproduced the
M-L-T spectral transition satisfactorily.
  
Attempts have been made to account for atmospheric dynamics 
in planetary atmospheres, whose models however cat not describe the convection 
cells and the resulting gravity waves \citep{Marley2007prpl.conf..733M,
Fortney2006ApJ...652..746F}.
However, local radiation hydrodynamics (RHD) models of the surface
layers of the solar convection have been very successful in
reproducing and analyzing the properties of the granulation
\citep{Nordlund1982A&A...107....1N}.
In the meantime, various groups have developed similar codes
to investigate the atmospheric flows on the sun and other stars
\citep{Steffen1989A&A...213..371S,
Asplund2000A&A...359..669A,
Skartlien2000ApJ...541..468S,
Stein2000SoPh..192...91S,
Gadun2000A&AS..146..267G,
Robinson2003MNRAS.340..923R,
Voegler2004A&A...421..755V}.
Amongst others, these models can describe self-consistently the mixing of
material beyond the classical boundaries of a convection zone, as
demonstrated for instance for main-sequence A-type
stars
\citep{Freytag1996A&A...313..497F} or for M dwarfs
\citep{Ludwig2002A&A...395...99L,
Ludwig2006A&A...459..599L}.
A treatment of dust within a 3D simulation of the envelope of an AGB star
was included by
\citet{Freytag2008A&A...483..571F}.

The aim of the current work is to extend the latter simulations into
the regime of brown dwarfs,
where dust clouds have a strong influence on the photospheric temperature structure,
and to quantify the overshoot from the surface convection zone
into the atmosphere.

%In Sect.~\ref{s:2Dsimulations} we present the numerical method
%and the setup of the models.
%Section~\ref{s:Results} describes the results of the
%2D simulations.
%Section~\ref{s:Discussion} contains our discussion
%and Section~\ref{s:Conclusions} our conclusions.

%#########################################################################################
\section{Simulations with CO5BOLD \label{s:2Dsimulations}}

%===============================================================================
\subsection{Numerical radiation hydrodynamics}

\begin{table*}[htb]
 \begin{center}
  \caption{Basic parameters of the RHD models (mostly 2D and only one 3D)
\label{t:modelparam}}
  \begin{tabular}{lrrrrrrrrr} \hline
model & p & $n_x \times n_z$ & $x \times z$ & $t_0 - t_1$ & $C_\mathrm{Cour}$ & $s_\mathrm{in}$ & $T_\mathrm{eff,sta}$ & $T_\mathrm{eff}$ & $T_\mathrm{opa}$ \\
  &   &   & km$\times$km & 10$^3$s &   & erg\,K$^{-1}$g$^{-1}$ & K & K & K \\
mt09g50mm00n10   & d & 400$\times$430 & 220$\times$ 94 &     90 -    130 & 0.30 & 0.671$\,10^9$ &  900 &  897 & 1000 \\
mt09g50mm00n11   & d & 400$\times$410 & 220$\times$ 90 &    240 -    275 & 0.30 & 0.671$\,10^9$ &  900 &  897 & 1000 \\
mt10g50mm00n03   & d & 400$\times$400 & 240$\times$ 96 &    140 -    195 & 0.27 & 0.683$\,10^9$ & 1000 & 1030 & 1000 \\
mt11g50mm00n05   & d & 400$\times$356 & 260$\times$ 92 &    245 -    285 & 0.30 & 0.690$\,10^9$ & 1100 & 1114 & 1800 \\
mt12g50mm00n01   & s & 400$\times$336 & 280$\times$ 94 &    110 -    190 & 0.40 & 0.701$\,10^9$ & 1200 & 1226 & 1800 \\
mt12g50mm00n07   & s & 400$\times$380 & 280$\times$106 &     80 -    150 & 0.40 & 0.701$\,10^9$ & 1200 & 1228 & 1800 \\
mt12g50mm00n10   & s & 400$\times$336 & 280$\times$ 94 &    270 -    370 & 0.37 & 0.701$\,10^9$ & 1200 & 1224 & 1800 \\
mt13g50mm00n01   & s & 400$\times$270 & 300$\times$100 &     80 -    190 & 0.40 & 0.711$\,10^9$ & 1300 & 1335 & 1800 \\
mt13g50mm00n02   & d & 400$\times$270 & 300$\times$100 &    130 -    200 & 0.40 & 0.711$\,10^9$ & 1300 & 1336 & 1800 \\
mt13g50mm00n03   & s & 400$\times$270 & 300$\times$100 &    250 -    390 & 0.35 & 0.711$\,10^9$ & 1300 & 1333 & 1800 \\
mt14g50mm00n01   & s & 400$\times$270 & 320$\times$106 &     40 -    140 & 0.40 & 0.719$\,10^9$ & 1400 & 1436 & 1800 \\
mt14g50mm00n02   & s & 400$\times$270 & 320$\times$106 &    140 -    280 & 0.40 & 0.719$\,10^9$ & 1400 & 1437 & 1800 \\
mt15g50mm00n01   & s & 400$\times$270 & 340$\times$113 &     50 -    140 & 0.40 & 0.728$\,10^9$ & 1500 & 1533 & 1800 \\
mt15g50mm00n02   & s & 400$\times$270 & 340$\times$113 &    150 -    190 & 0.40 & 0.728$\,10^9$ & 1500 & 1533 & 1800 \\
mt15g50mm00n03   & s & 400$\times$270 & 340$\times$113 &    260 -    380 & 0.30 & 0.728$\,10^9$ & 1500 & 1533 & 1800 \\
mt15g50mm00n04   & s & 300$\times$270 & 340$\times$113 &     40 -    150 & 0.30 & 0.728$\,10^9$ & 1500 & 1532 & 1800 \\
mt15g50mm00n06   & s & 300$^2$$\times$270 & 340$^2$$\times$113 &     10 -     15 & 0.30 & 0.728$\,10^9$ & 1500 & 1532 & 1800 \\
mt15g50mm00n07   & s & 400$\times$300 & 340$\times$127 &     70 -    150 & 0.30 & 0.728$\,10^9$ & 1500 & 1533 & 1800 \\
mt16g50mm00n06   & s & 400$\times$366 & 352$\times$128 &     20 -    190 & 0.40 & 0.737$\,10^9$ & 1600 & 1648 & 1800 \\
mt17g50mm00n02   & s & 400$\times$352 & 380$\times$133 &     40 -    190 & 0.40 & 0.747$\,10^9$ & 1700 & 1757 & 1800 \\
mt18g50mm00n07   & s & 400$\times$343 & 400$\times$137 &     40 -    190 & 0.40 & 0.756$\,10^9$ & 1800 & 1858 & 1800 \\
mt19g50mm00n02   & s & 400$\times$334 & 420$\times$140 &     40 -     80 & 0.40 & 0.765$\,10^9$ & 1900 & 1953 & 1800 \\
mt19g50mm00n03   & s & 400$\times$334 & 420$\times$140 &    120 -    290 & 0.30 & 0.765$\,10^9$ & 1900 & 1953 & 1800 \\
mt20g50mm00n05   & s & 400$\times$362 & 380$\times$137 &     40 -    210 & 0.20 & 0.775$\,10^9$ & 2000 & 2052 & 1800 \\
mt22g50mm00n05   & s & 400$\times$351 & 400$\times$140 &     70 -    150 & 0.20 & 0.798$\,10^9$ & 2200 & 2247 & 1800 \\
mt24g50mm00n01   & s & 400$\times$344 & 420$\times$144 &     70 -    190 & 0.20 & 0.825$\,10^9$ & 2400 & 2426 & 1800 \\
mt26g50mm00n01   & s & 400$\times$353 & 420$\times$148 &     70 -    140 & 0.20 & 0.859$\,10^9$ & 2600 & 2597 & 1800 \\
mt26g50mm00n02   & s & 400$\times$353 & 420$\times$148 &     70 -    140 & 0.20 & 0.859$\,10^9$ & 2600 & 2611 & 2800 \\
mt28g50mm00n01   & s & 400$\times$357 & 440$\times$157 &     70 -    140 & 0.20 & 0.905$\,10^9$ & 2800 & 2801 & 2800 \\
mt28g50mm00n02   & s & 400$\times$380 & 440$\times$167 &     40 -    100 & 0.20 & 0.905$\,10^9$ & 2800 & 2801 & 2800 \\
\hline
  \end{tabular} \\
 \end{center}
The columns show
model name,
numerical precision (single or double),
horizontal $\times$ vertical resolution,
horizontal $\times$ vertical size [km$\times$km],
time span used for averaging [10$^3$s],
Courant number,
entropy of the material in the deeper layers [erg\,K$^{-1}$g$^{-1}$],
effective temperature of the PHOENIX model used for the start file [K],
effective temperature of the RHD model versus the average time span [K], and
effective temperature of the reference atmosphere used for the opacity table [K].
\end{table*}

We computed a sequence of
2D RHD models for a gravity of $10^5$\,cm\,s$^{-2}$ (log\,$g$=5)
and a range of effective temperatures from 900\,K to 2800\,K.
The models have about 400$\times$300 grid points
(see Table~\ref{t:modelparam} for details).
Most of them are restricted to two dimensions because we are unable
to cover the prohibitively long sedimentation and mixing timescales in 3D:
a 2D simulation in itself takes about one to three CPU-months to complete.
However, a shorter run covering only several dynamical timescales and not
trying to cover the longer mixing timescales is feasible in 3D (mt15g50mm00n06).

For this purpose, we used the multi-D RHD code
CO5BOLD\footnote{CO5BOLD User Manual: \href{http://www.astro.uu.se/~bf/co5bold_main.html}{http://www.astro.uu.se/$\tilde{\,\,}$bf/co5bold\_main.html}}
\citep{Freytag2002AN....323..213F,
Wedemeyer2004A&A...414.1121W}
%\citep{Freytag2004CO5BOLD-Manual}
in its local box setup to calculate time-dependent atmosphere models,
including the very top layers of the convection zone.
To realize this project, we implemented a dust model (see below)
as well as dust and low temperature gas opacities.

The code solves the coupled equations of compressible hydrodynamics
and non-local radiation transport on a Cartesian grid with a
time-explicit scheme.
The tabulated equation of state accounts for the ionization of hydrogen and helium,
and the formation of H$_2$~molecules.
The 1D hydrodynamics fluxes are computed with an approximate Riemann solver of Roe-type.
Because the conditions in the cool objects are almost incompressible,
the fluxes are combined non-split,
i.e.,\ the fluxes in both the vertical and horizontal directions are computed
from the same state (and not after each other) and their contributions are added.
In this way, the generation of spurious pressure waves is avoided,
which may be produced by a split scheme
in regions with large gradients but small divergence in the mass flux.

%-------------------------------------------------------------------------------
\subsubsection{Dust model}

To account for the presence of dust particles, we added terms in
the modules for hydrodynamics, radiation transport, source terms, and
in handling of boundary conditions. It is impossible to account
for all microphysical processes that might play a role in dust
formation
\citep{Helling2001A&A...376..194H, Woitke2003A&A...399..297W}
in current time-dependent multi-dimensional simulations.
We instead chose a treatment of dust that includes only
the most important physical processes.
The scheme is based on a simplified version of the dust model used in
\cite{Hoefner2003A&A...399..589H}.
We use a single density field to describe the mass density of dust particles
and qno5h34 for the monomers (gas constituents),
instead of four for the dust and none for the monomers as in
\cite{Hoefner2003A&A...399..589H} and
\cite{Freytag2008A&A...483..571F}.
Therefore, the ratio of the sum of dust and
monomer densities to the gas density is allowed to change,
in contrast to the dust description by \cite{Hoefner2003A&A...399..589H}.
Instead of modeling the nucleation and the detailed evolution of the
number of grains, we assume a constant ratio of seeds (dust nuclei)
to total number of monomers (in grains or free) per cell.
If all the material in a grid cell were to be condensed into dust, the
grains would have the maximum radius $r_\mathrm{d,max}$, which we
have set to a typical value of 1\,$\mu$m. This is close to the
typical particle sizes found for the optically thick part of the
cloud deck in solar-metallicity brown dwarfs according to the
\texttt{DRIFT-PHOENIX} models of \citet{Witte2009A&A...506.1367W} 
and according to our own \texttt{PHOENIX} BT-Settl calculations.
In both models, particle sizes are determined by a balance between settling speed
and turbulent upmixing according to a basic convective overshoot model (cf.\
\citealt{Helling2008MNRAS.391.1854H} for a comparison), and are thus
in general height-dependent, reaching up to several $\mu$m in the
deepest cloud layers. For the present models, the value chosen here 
should allow a reasonable estimate of the dust opacity in the denser
parts of the cloud deck.
The radius $r_\mathrm{d}$ of dust grains for given dust mass density
$\rho_\mathrm{d}$ and monomer mass density $\rho_\mathrm{m}$ is
computed from
\begin{equation}\label{eq:rdust}
r_\mathrm{d} = r_\mathrm{d,max} \, [ \rho_\mathrm{d} / ( \rho_\mathrm{d} + \rho_\mathrm{m} )]^{1/3} \enspace .
\end{equation}
Condensation and
evaporation are modeled as in \cite{Hoefner2003A&A...399..589H},
parameters and saturation vapor curve adapted to forsterite.

In the hydrodynamics module, monomers and dust densities are advected
with the gas density.
However,
according to 
the terminal velocities given by the low-Reynolds-number case of Eq.~(19) in
\citet{Rossow1978Icar...36....1R},
a settling speed
is added to the vertical advection velocity of dust grains,
assuming instantaneous equilibrium
between gravitational and viscous forces that act onto the grains.

One problem with modeling the dynamics of dust clouds is the span in
timescales (short for dust formation and the wave period, long for
dust settling and thermal relaxation) and spatial scales (small-scale
dust clouds and possible global flows caused by rapid rotation).  This is
quite similar to simulations of weather patterns on Earth, where global
wind systems and local cloud formation interact.

Another problem is the poorly known complex microphysics:
a complicated chemical network of molecules
with space- and time-dependent abundances
can form dust by means of various processes,
producing grains with different structures.
The dynamical behavior and optical properties both depend on the grain type.
Furthermore, depletion leads to a change in the gas composition that affects
the equation of state and gas opacities.
The current dust model in CO5BOLD is designed to reproduce the essential processes,
but cannot account for all details that might possibly play a role.

%-------------------------------------------------------------------------------
\subsubsection{Equation of state and opacities}

The equation of state accounts for the ionization of hydrogen and
helium, and the formation of molecular hydrogen. CO5BOLD can deal with
the effects of ionization but not with an element composition
that depends on space and time. Therefore, the depletion of elements is
ignored for the equation of state: the formation of molecules has only
a minor effect on e.g.,\ the heat capacity as long as hydrogen exists in
the form of $H_2$. However, molecules play a major role for the
opacity, and the formation of molecules depends both on the abundance and
depletion of elements. To take this into account, we derived the
CO5BOLD gas phase opacities from monochromatic opacity tables,
$\kappa(T,P,\nu)$, generated from detailed radiation transfer
calculations with the general stellar atmosphere code
\texttt{PHOENIX} \citep{Hauschildt1997ApJ...483..390H}.
%We assume dust formation in equilibrium with the gas phase:
We assume full sedimentation of dust from the gas phase:
the removal of condensable material from the gas phase
is considered assuming a solar elemental composition in full phase
equilibrium at each temperature and pressure point
\citep[see][]{Allard2001ApJ...556..357A, Ferguson2005ApJ...623..585F}.
This approximation is close to the conditions prevailing in:
i) the lower atmospheric layers that are too hot for dust condensation,
ii) the uppermost layers where the gravitational settling depletion is
partially compensated by dynamical upwelling of monomers, and
iii) in the cloud-forming layers as confirmed by observations as stated
above. The monochromatic gas opacity table was averaged
into 5~bins to minimize the computing time but retain the radiative
equilibrium properties of the gas.

In contrast to the sophisticated treatment of the gas opacities,
we use a simple formula for the dust opacities,
which assumes that the large particle limit is valid for all grain sizes
and treats scattering as true absorption.
The dust opacity [cm$^{-1}$] is
\begin{equation}
 \kappa_\mathrm{d} = 3/(2 \, \rho_\mathrm{d,material} \, r_\mathrm{d,max}) \,\,
                    [ \rho_\mathrm{d}^2  ( \rho_\mathrm{d} + \rho_\mathrm{m} )]^{1/3}  \enspace ,
\end{equation}
computed dynamically from the quantities as described for Eq.~(\ref{eq:rdust})
in each cell of the simulated atmosphere and added to the gas opacity.
We concentrate on forsterite grains (Mg$_2$SiO$_4$, 3.3\,g/cm$^3$) that
are relatively abundant and provide the greatest contribution to the total
dust opacities.

%-------------------------------------------------------------------------------
\subsubsection{Boundary conditions}

The side boundaries of the computational domain are periodic.
Usually, an open top is used together with an open bottom boundary
conditions for local models that comprise part of a deep convection
zone.  However, closed boundaries keep the amount of dust and
condensable material constant within the computational domain.
We therefore used closed boundaries (top and bottom) for all brown
dwarf models, although the stellar convection zone should extend to
the center of the star. To keep the entropy close to a prescribed
value, the internal energy is adjusted for a few grid layers (10~km
height) at the bottom of the model. This mechanism acts as an energy
source and replenishes the radiative energy losses through the top of
the model. This parameter (the value of the entropy plateau
$s_\mathrm{in}$ in the deep convective layers) controls the effective
temperature and is taken from the start-up {\it Dusty} models. Moreover, a drag
force dampens downdrafts in these bottom layers.

The top boundary is closed as well, partly to keep material inside.
It has a damping zone of about 8 grid points where a strong drag force is
applied. Damping at an open boundary did not appear sufficient to
keep gravity waves with moderate Mach number (with peak values close to 1)
from achieving additional growth to implausibly large amplitudes.

%-------------------------------------------------------------------------------
\subsubsection{Initial conditions}

The thermal
structure of a start model is based on a classical
1D stationary stellar atmosphere model produced with \texttt{PHOENIX} assuming
hydrostatic equilibrium and radiative plus convective
(using the Mixing-Length Theory,  \citealt{BohmVitense1958ZA.....46..108B})
flux equilibrium. We preferred the dust-rich {\it Dusty} over the
dust-free {\it Cond} models even for lower temperatures where the
{\it Cond} models represent dust-free photospheres, because the resulting
effective temperature of the CO5BOLD models agrees very well with the
effective temperature of the {\it Dusty} models (CO5BOLD and \texttt{PHOENIX}
models have per construction the same entropy in the deeper layers --
not necessarily the same effective temperature).  
We interpolated the {\it Dusty}
grid points to a finer grid with small or no variation in the grid
spacing. To allow sufficient volume for the surface granules to form, we
added several points at the bottom by integrating a hydrostatic
stratification with constant entropy, taken from the bottom point of
the {\it Dusty} model. At the top, we attached a few points with the
internal energy value of the top point in the {\it Dusty} model,
to maintain a sufficient distance between the top of the cloud layers and the top
boundary of the computational box. 

We enlarged the model in one
horizontal dimension to 400~points, and imposed small random
velocity fluctuations as seeds for convective instability.  Initially,
we set a constant fraction of the monomers plus dust mass density
divided by the gas density, but which we reduced somewhat
empirically in the uppermost layers to account for the partial
depletion of material. The relative amount of material in the monomer
bin is then determined by the saturation pressure of forsterite.
Although the {\it Dusty} models assume hydrostatic
equilibrium, there are small deviations from exact numerical
equilibrium in the initial CO5BOLD models.
These cause unwanted plane-parallel oscillations that we suppressed
by a drag force in the initial phase of each simulation.
To dampen these, we
applied a strong drag force acting only on plane-parallel motions
within the first 100\,sec. In the following 9900\,sec, we reduced the
drag force to remove remaining plane-parallel residuals.  For the remainder
of the run (including the interval where we take averages from), we
still have a very small but non-zero drag force that dampens plane-parallel
vertical and horizontal motions on a timescale of 15\,000\,sec to
suppress some modes that grew in early models over very long timescales.

%#########################################################################################
\section{Results of the simulations \label{s:Results}}

% ..............................................................................
\begin{figure*}
\centering
\includegraphics[]{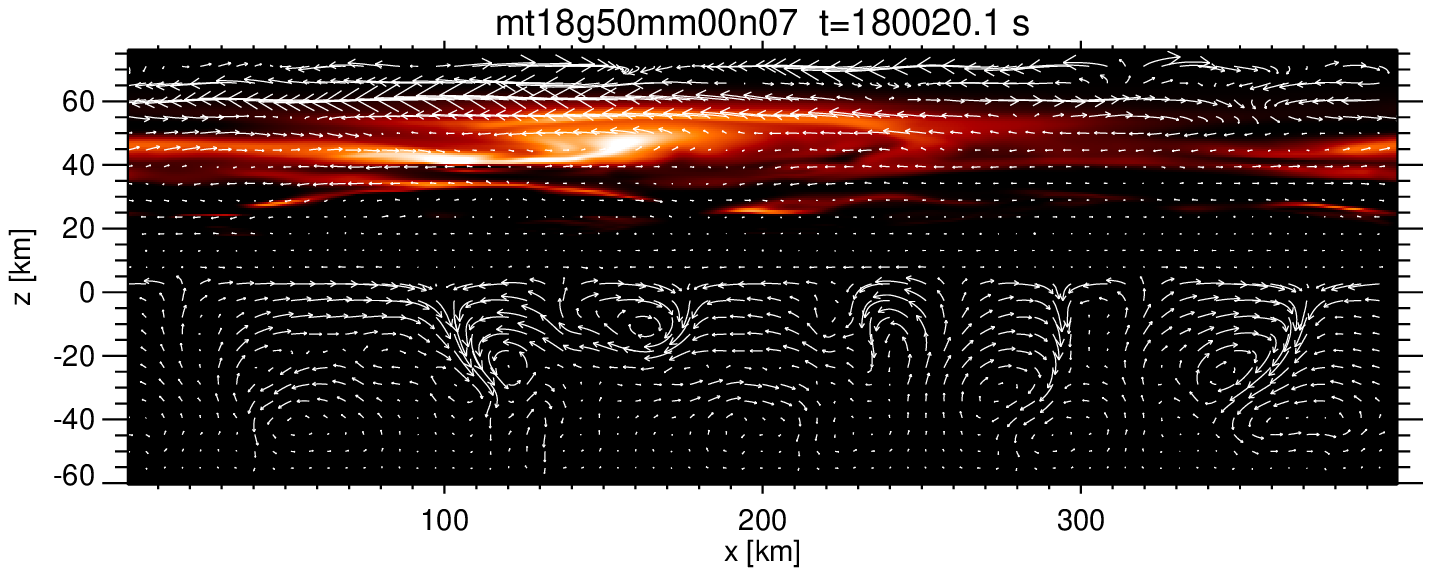}
\includegraphics[]{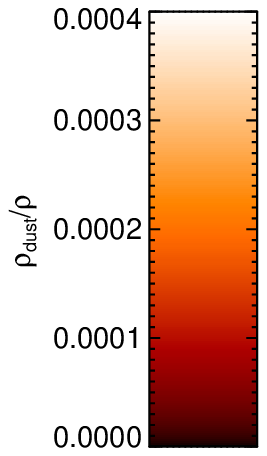}
\caption{This snapshot
from a brown dwarf simulation
with $T_{\rm eff}$=1858\,K, $\log\,g$=5
shows the velocity field as pseudo-streamlines, color-coded according to the dust concentration.
The flow in the lower part is due to the surface granulation of the stellar convection zone.
The top is dominated by gravity waves.
}
\label{f:aabd1_2dslice_mt18g50mm00n07_vel_qucallm1orho}
\end{figure*}

% ..............................................................................
\begin{figure}
\centering
\includegraphics[]{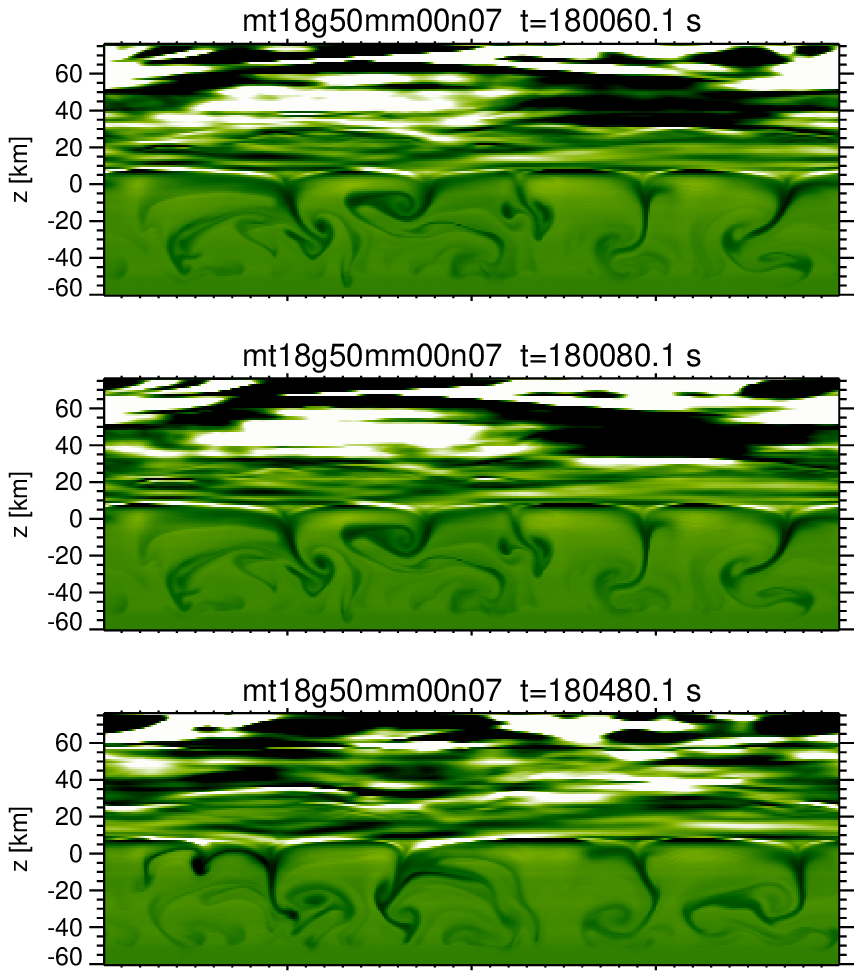}
\includegraphics[]{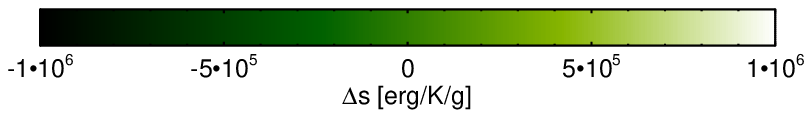}
\caption{Three snapshots of the entropy fluctuations
(entropy with horizontal average removed)
of a brown dwarf model mt18g50mm00n07 with
$T_{\rm eff}$=1800\,K and $\log g$=5.}
\label{f:aabd1_2dslice_mt18g50mm00n07_s-ms}
\end{figure}

% ..............................................................................
\begin{figure}
\centering
\includegraphics[]{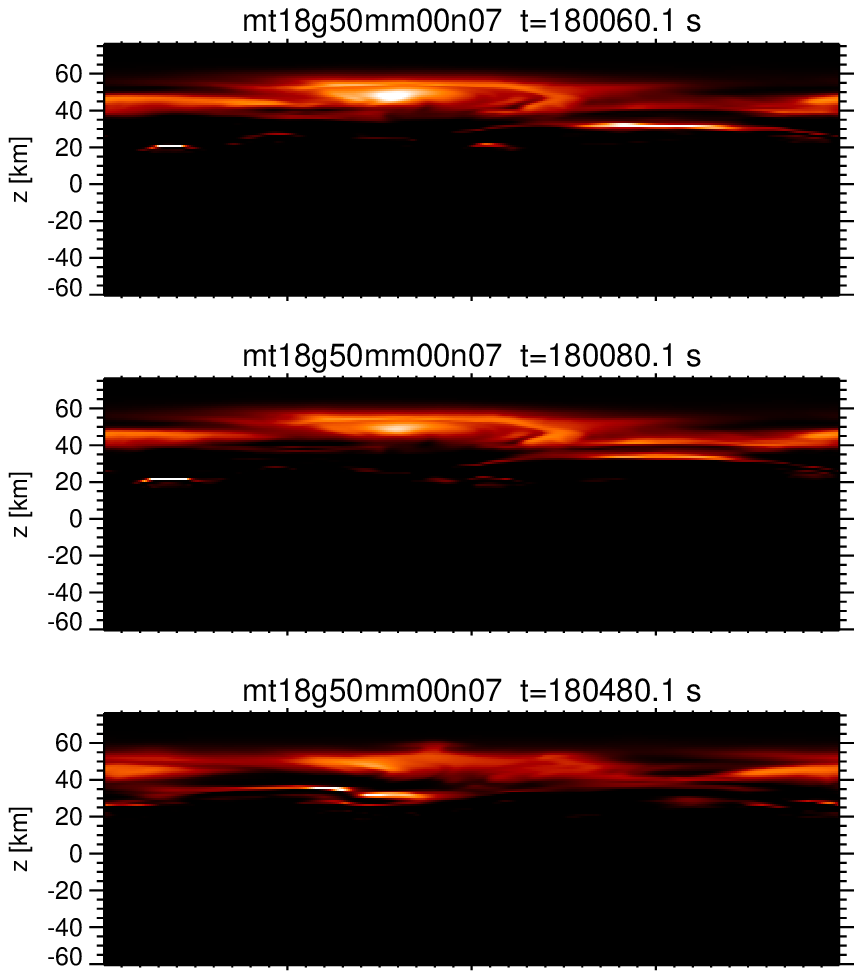}
\includegraphics[]{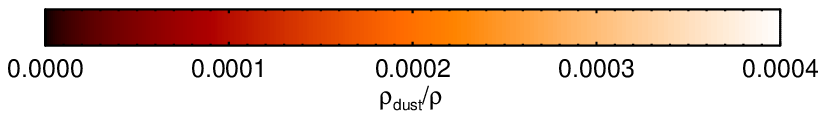}
\caption{Three snapshots of the dust concentration of a brown dwarf model mt18g50mm00n07 with
$T_{\rm eff}$=1800\,K and $\log g$=5.}
\label{f:aabd1_2dslice_mt18g50mm00n07_qucallm1orho}
\end{figure}

% ..............................................................................
\begin{figure}
\centering
\includegraphics[]{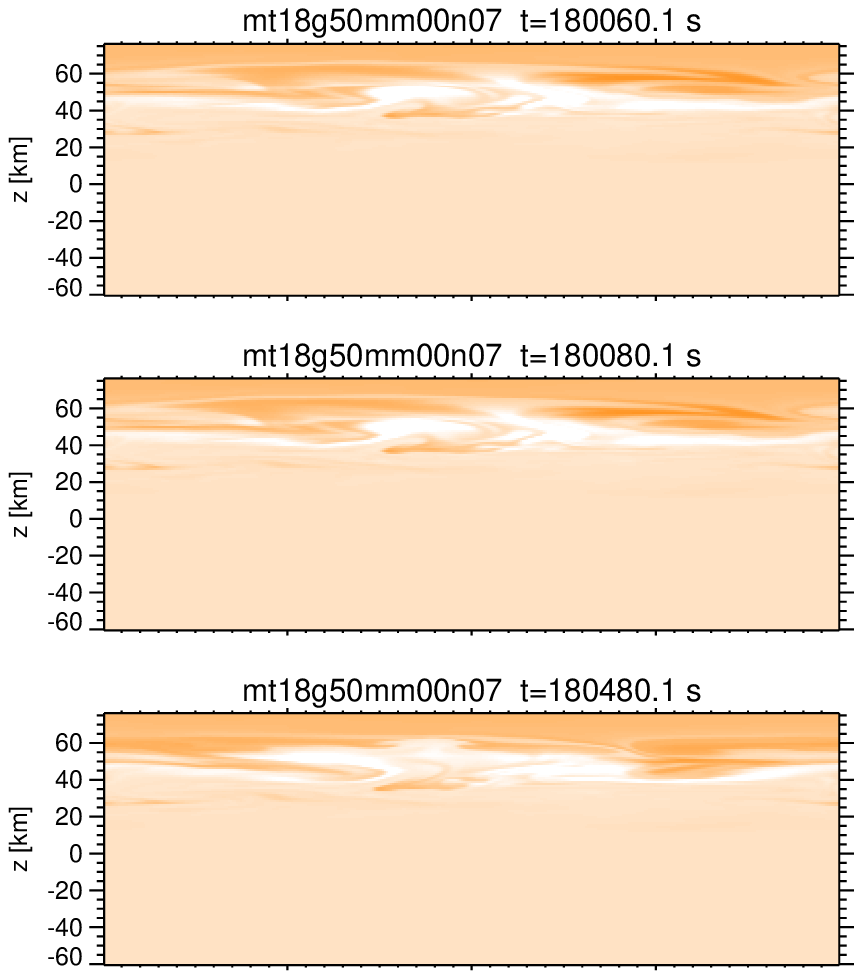}
\includegraphics[]{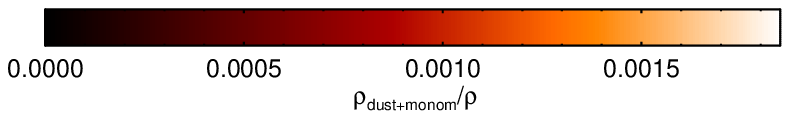}
\caption{Three snapshots of the concentration of dust+monomers
of a brown dwarf model mt18g50mm00n07 with
$T_{\rm eff}$=1800\,K and $\log g$=5.}
\label{f:aabd1_2dslice_mt18g50mm00n07_qucallorho}
\end{figure}

% ..............................................................................
\begin{figure*}
\centering
\includegraphics{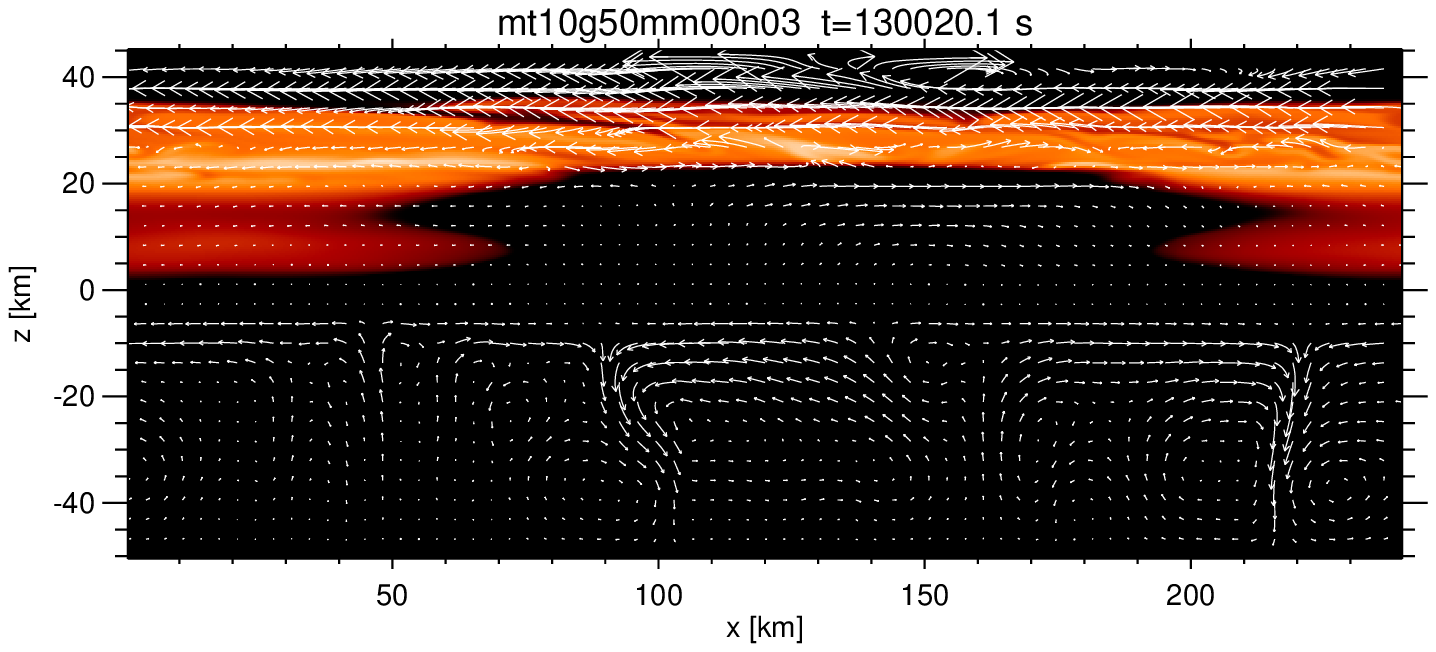}
\includegraphics{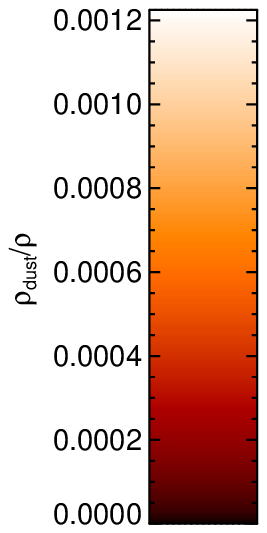}
\caption{As Fig.~\ref{f:aabd1_2dslice_mt18g50mm00n07_vel_qucallm1orho}
for a brown dwarf model (mt10g50mm00n03) with
$T_{\rm eff}$=1000\,K and $\log g$=5.}
\label{f:aabd1_2dslice_mt10g50mm00n03_vel_qucallm1orho}
\end{figure*}

% ..............................................................................
\begin{figure}
\centering
\includegraphics[]{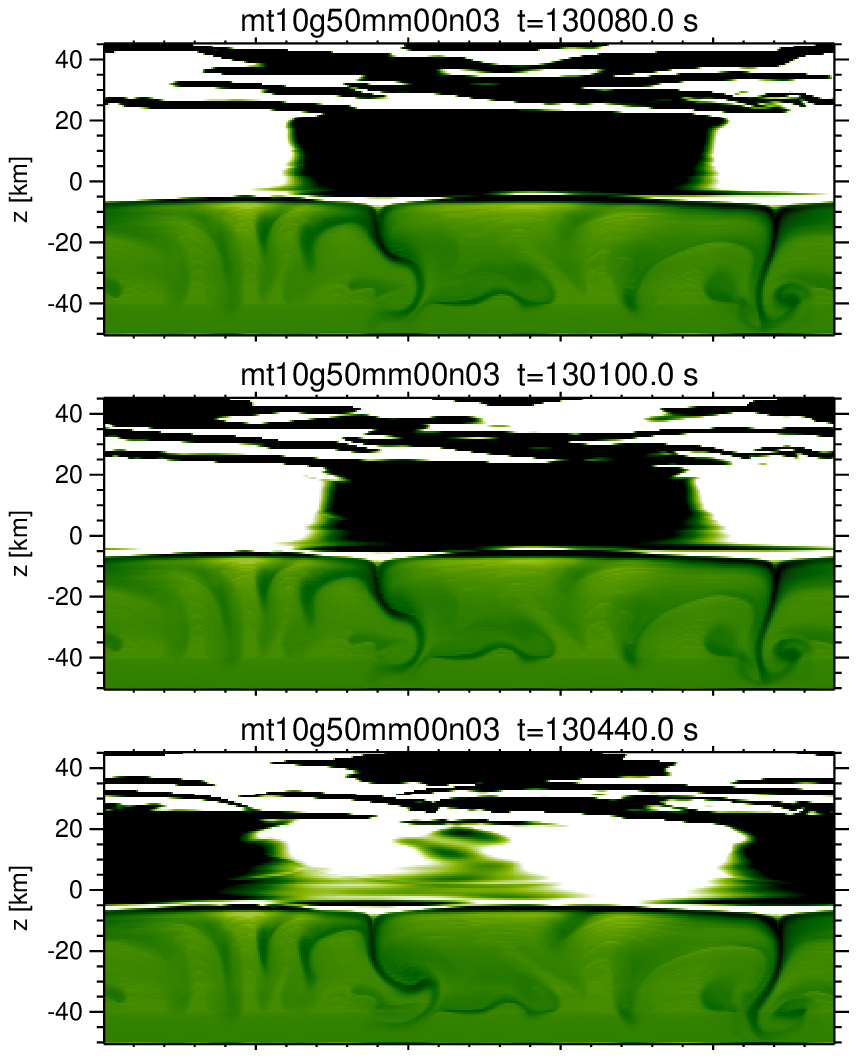}
\includegraphics[]{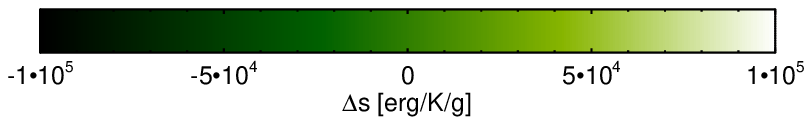}
\caption{Three snapshots of the entropy fluctuations of a brown dwarf model mt10g50mm00n03 with
$T_{\rm eff}$=1000\,K and $\log g$=5.}
\label{f:aabd1_2dslice_mt10g50mm00n03_s-ms}
\end{figure}

% ..............................................................................
\begin{figure}
\centering
\includegraphics[]{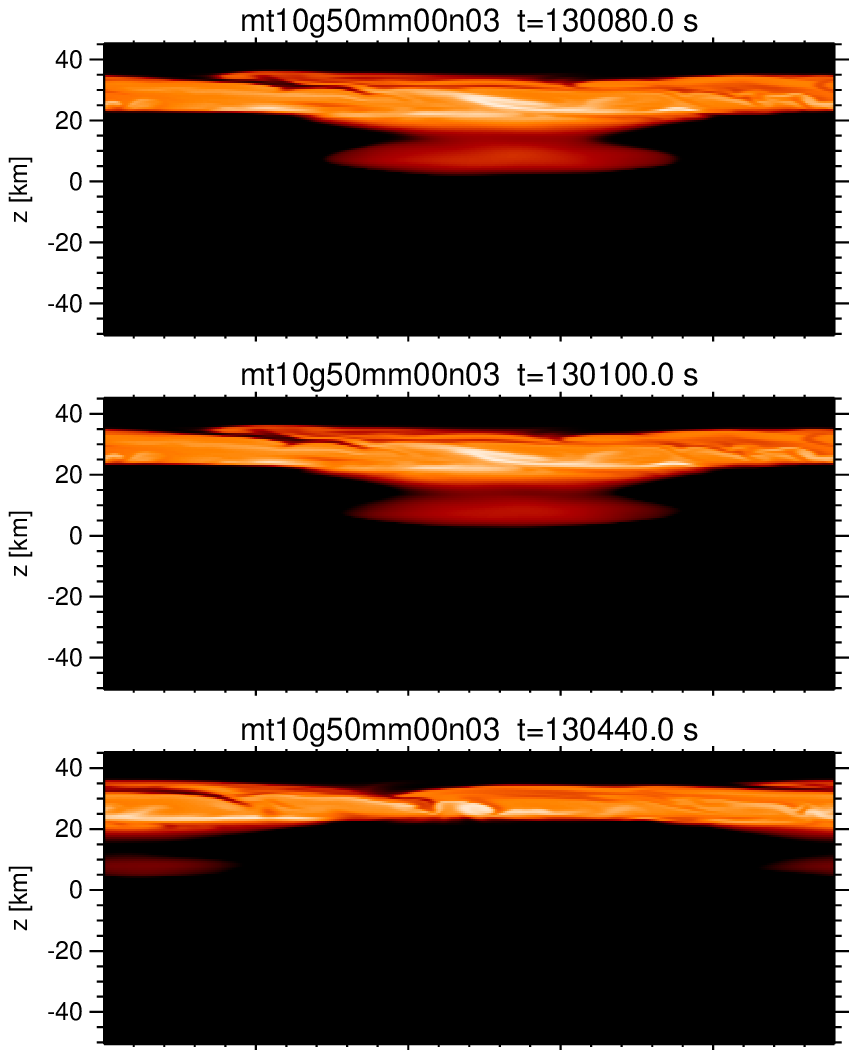}
\includegraphics[]{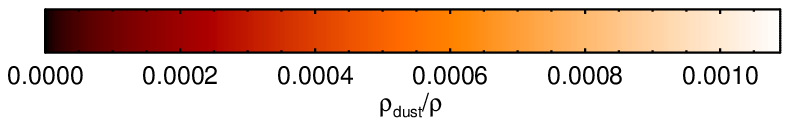}
\caption{Three snapshots of the dust concentration of a brown dwarf model mt10g50mm00n03 with
$T_{\rm eff}$=1000\,K and $\log g$=5.}
\label{f:aabd1_2dslice_mt10g50mm00n03_qucallm1orho}
\end{figure}

% ..............................................................................
\begin{figure}
\centering
\includegraphics[]{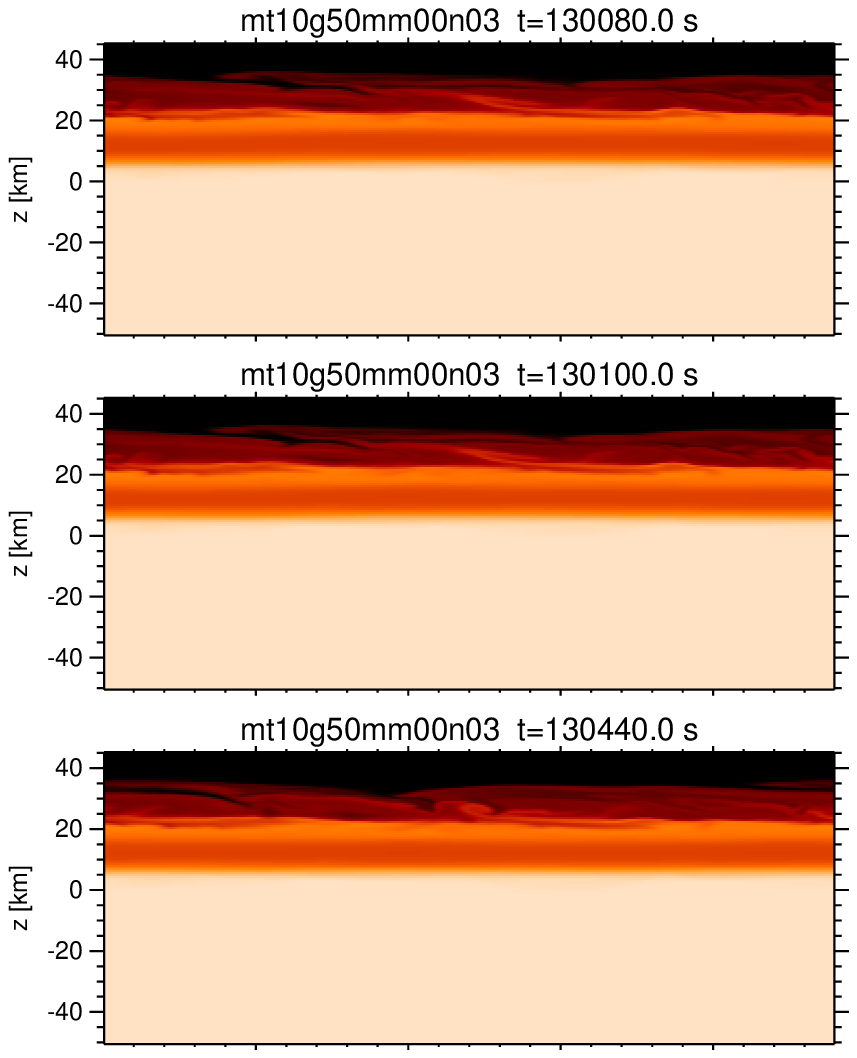}
\includegraphics[]{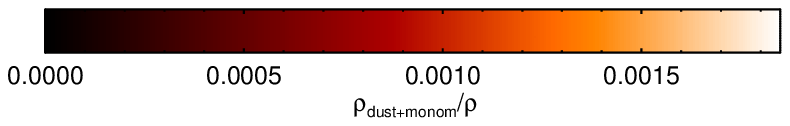}
\caption{Three snapshots of the concentration of dust+monomers
of a brown dwarf model mt10g50mm00n03 with
$T_{\rm eff}$=1000\,K and $\log g$=5.}
\label{f:aabd1_2dslice_mt10g50mm00n03_qucallorho}
\end{figure}

% ..............................................................................
% ... aabd1_seqavgstat_log10_v3rms_log10_P_plot.pro ...
\begin{figure*}
\centering
\includegraphics[]{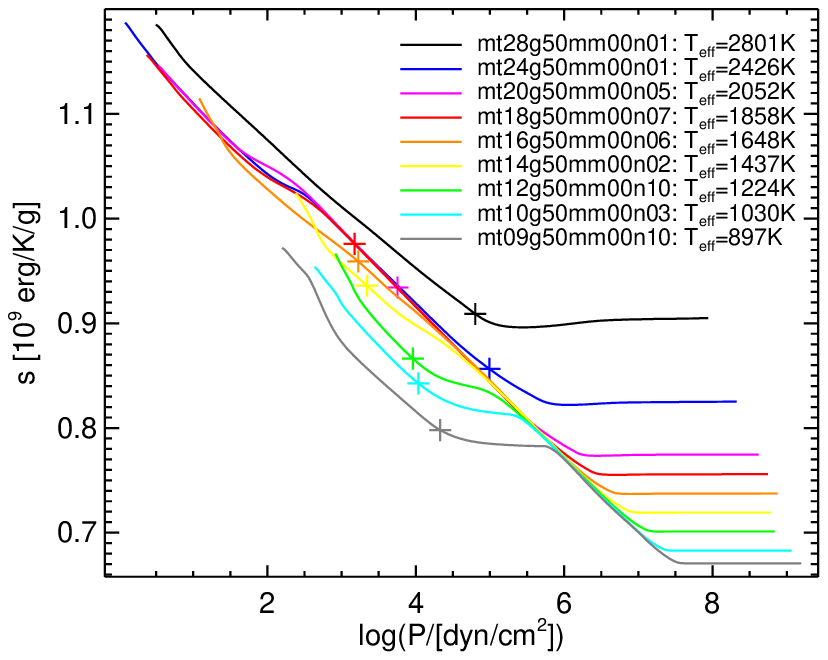}
\includegraphics[]{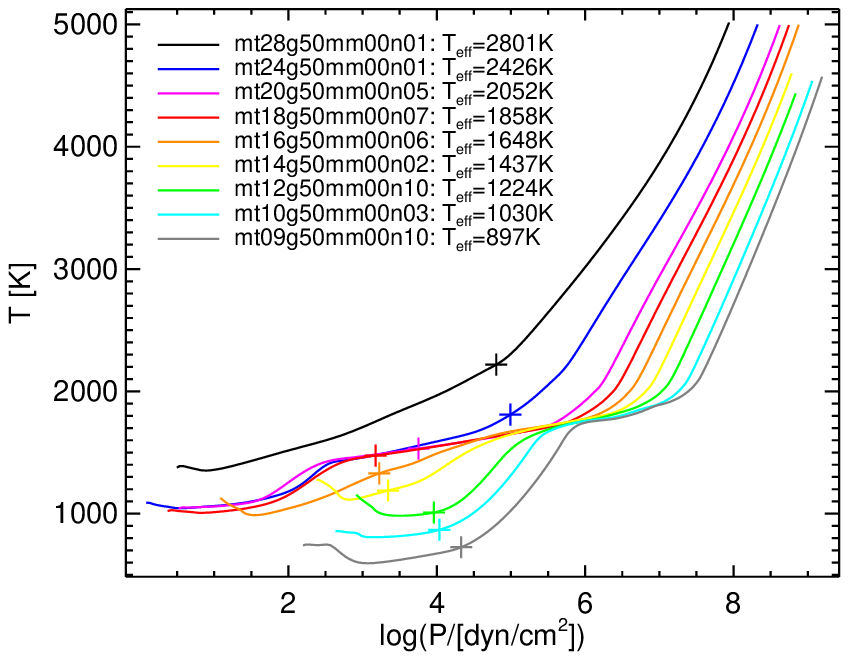}
\includegraphics[]{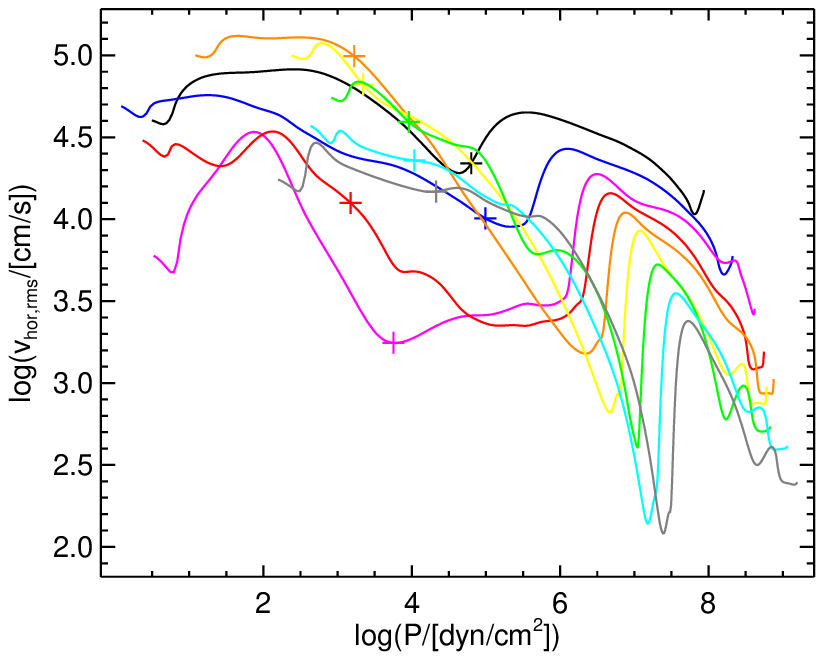}
\includegraphics[]{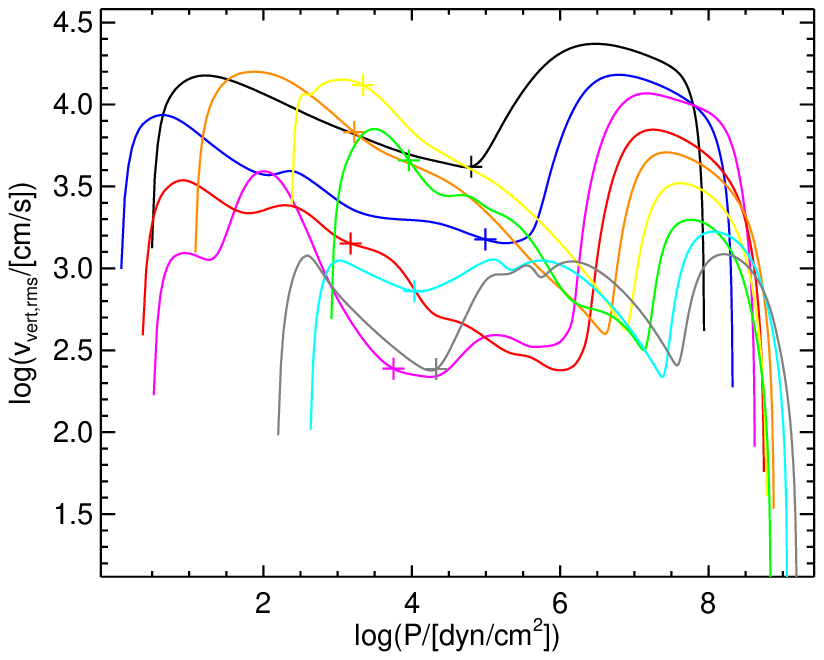}
\includegraphics[]{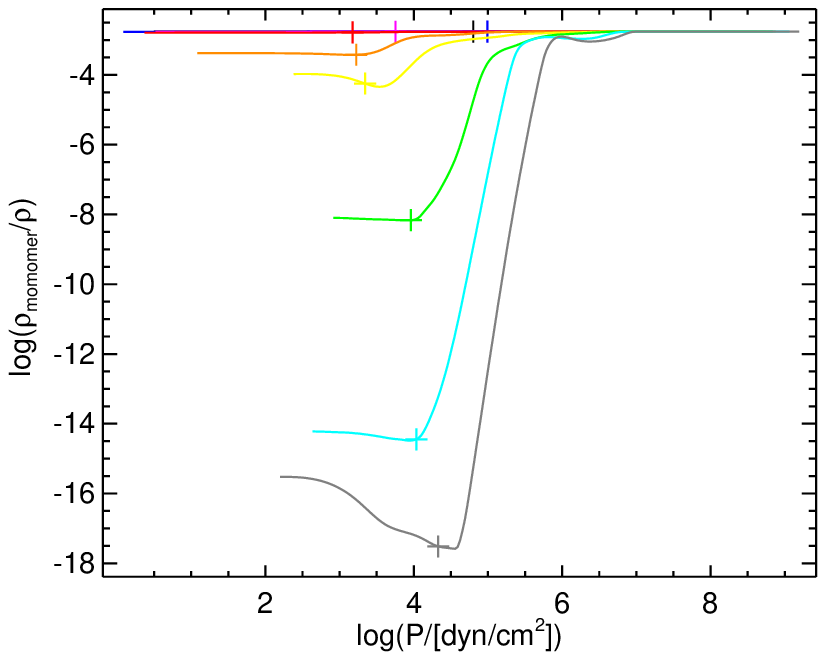}
\includegraphics[]{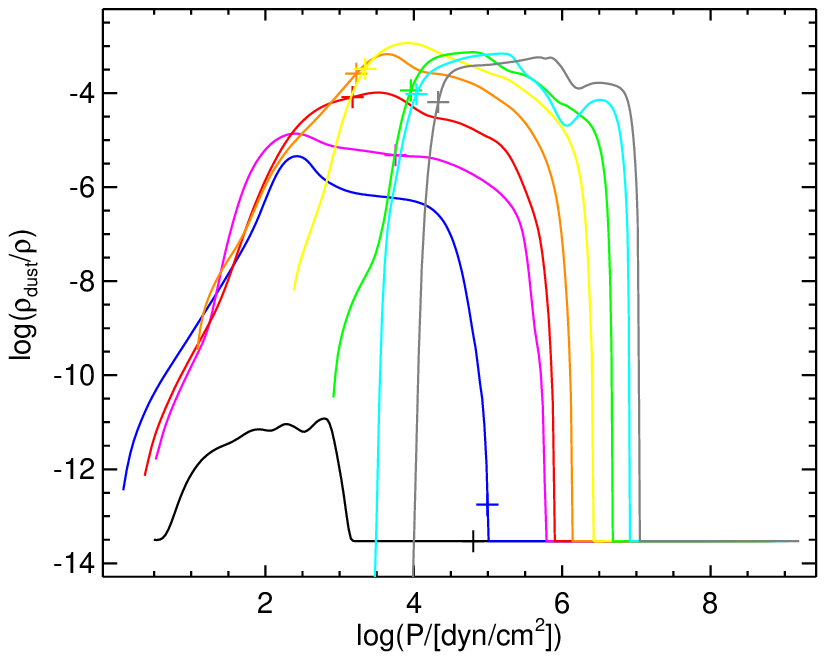}
\caption{
Various averaged quantities versus logarithm of pressure:
Top left: mean entropy
for various effective temperatures and $\log g$=5.
The plus signs mark the layers with Rosseland optical depth $10^{-2}$.
The legend is the same in all panels.
Top right: mean temperature.
Middle left: logarithm of rms horizontal velocity.
Middle right: logarithm of rms vertical velocity.
Bottom left: logarithm of monomer concentration.
Bottom right: logarithm of dust concentration.
}
\label{f:aabd_seqavgstat_q_log10_P}
\end{figure*}

%===============================================================================
\subsection{Convection and timescales}

The quiet solar surface -- far away from sun spots -- is characterized
by a mottled pattern of bright hot rising areas
surrounded by dark lanes of cool downflowing material -- the so-called granulation
at the top of the solar convection zone.
Because of their higher surface gravity and lower effective temperature
(log\,$g$=5,  $T_\mathrm{eff}$$\approx$1800\,K),
granules on brown dwarfs are about a factor of 10
smaller than their counterparts on the sun
(log\,$g$=4.44, $T_\mathrm{eff}$=5775\,K).
Only tiny velocities are required to transport the energy flux through
the convection zone, resulting in nearly incompressible
low-Mach-number flows
(with a maximum vertical rms value ranging
from about 0.06 at $T_\mathrm{eff}$=2800\,K
to about 0.003 at $T_\mathrm{eff}$=900\,K).
Accordingly, only small-amplitude pressure
waves are present in the brown dwarf models.
For comparison,
the acoustic timescale (for an up-down-up wave travel)
is about 90\,s for a 2800\,K model
and about 60\,s at the cool end of our sequence.
The Brunt-V{\"a}is{\"a}l{\"a} period in the stable layers is about 25\,s,
while the convective growth time in the unstable layers increases with decreasing effective
temperature from around 40\,s to about 300\,s (minimum values at the ``most unstable layers'').
The free-fall timescale (to drop one pressure scale height)
is around 3\,s in the photosphere.

To check the transition from the starting conditions to a quasi-stationary state,
we consider time sequences of spatial-averaged quantities,
such as temperature, rms velocities, and dust concentration.
Starting from random initial fluctuations, the onset of
convection takes a few 100~seconds,
longer at lower effective temperatures.
A statistically stable pattern develops after a few 1000~seconds.
Wave amplitudes relax on somewhat longer timescales.
In contrast, the thermal relaxation time --
particularly of the deeper convective layers -- is much longer. However, the thermal
structure of these layers -- an adiabat -- remains essentially the same as
in the initial model due to our choice of treatment of the lower
boundary (keeping the entropy constant instead of imposing a certain flux).
In this way, there is no need to cover the complete thermal relaxation time.
The longest timescale to be covered is the relaxation time of the dust concentration
that has an effect onto the temperature structure.
Therefore, each simulation covers a few days of stellar time.
The hydrodynamic time step is about 0.03\,s
and because of the relatively long radiative relaxation time, we perform multiple (typically 6)
hydrodynamical sub-steps per radiation transport step.

Snapshots from our atmosphere simulations are presented
in Figs.~\ref{f:aabd1_2dslice_mt18g50mm00n07_vel_qucallm1orho}
through \ref{f:aabd1_2dslice_mt10g50mm00n03_qucallorho}
while the complete videos are provided as supporting
material\footnote{\href{http://phoenix.ens-lyon.fr/papers/FreytagEtAl2009/}{http://phoenix.ens-lyon.fr/papers/FreytagEtAl2009/}}.
Figures~\ref{f:aabd1_2dslice_mt18g50mm00n07_vel_qucallm1orho}
and~\ref{f:aabd1_2dslice_mt10g50mm00n03_vel_qucallm1orho}
use pseudo-streamlines to visualize the flow field.
Figures~\ref{f:aabd1_2dslice_mt18g50mm00n07_s-ms} (for a 1800\,K model)
and \ref{f:aabd1_2dslice_mt10g50mm00n03_s-ms}
(for a 1000\,K model)
display sequences of the typical granulation pattern,
cool downdrafts occurring in a warmer environment.
It is clearly separated from the atmosphere in the upper half of the box
that shows inhomogeneities induced by gravity waves.
The downdrafts are relatively narrower than in solar granulation
\citep{Ludwig2002A&A...395...99L, Ludwig2006A&A...459..599L}.
In the image sequences, the first pair is 20\,s apart
whereas the last snapshot is taken several minutes later.

The entropy profiles -- averaged horizontally over constant height and in time ---
in Fig.~\ref{f:aabd_seqavgstat_q_log10_P}~(top left)
show a strong increase
in the upper atmosphere, with only a minor 
drop at the top of the convection zone,
and an almost flat distribution inside the convection zone.
This is indicative of very efficient convection resembling typical conditions in the 
stellar interior.

%===============================================================================
\subsection{Exponential overshoot}

The typical magnitude of the velocity fields can be inferred
from the plot of the rms of the vertical velocity versus pressure
for various effective temperatures
in Fig.~\ref{f:aabd_seqavgstat_q_log10_P} (middle panels).
The convective velocities fall significantly from the peak value
inside the convection zone (on the right)
until the top of the unstable layers, and even further into the overshoot region.
The scale height of
exponentially decreasing overshoot velocities
\citep{Freytag1996A&A...313..497F, Ludwig2006A&A...459..599L}
is so small that they do not induce
significant mixing in the cloud layers about two pressure scale heights further up.
Nevertheless, they are able
to mix material across the boundary between stable and unstable layers.

The bottom right panel in Fig.~\ref{f:aabd_seqavgstat_q_Teff} shows
the relatively large scale height of the convective velocity at high
effective temperatures.
This extended overshoot may play a role in the replenishment of dust material.
However, the overshoot scale height decreases rapidly
with $T_\mathrm{eff}$ and remains small ($H_v$$\approx$$0.28\,H_p$) from about 2200\,K on,
indicating that this type of overshoot is insignificant
for material mixing within the forsterite cloud layers.
On the other hand, for dust types that form at slightly higher temperatures
(around 2000\,K) as discussed e.g.,\ in \cite{Helling2004A&A...423..657H}
the mixing caused by convective overshoot can play a role.

%===============================================================================
\subsection{Gravity waves}

It is instead gravity waves that dominate the mixing of the atmospheric layers
(the upper half of the models in
Figs.~\ref{f:aabd1_2dslice_mt18g50mm00n07_vel_qucallm1orho}
to~\ref{f:aabd1_2dslice_mt10g50mm00n03_qucallorho})
with periods of about 30 to 100 seconds and
amplitudes that increase with height
(Fig.~\ref{f:aabd_seqavgstat_q_log10_P}).
Most prominent is the fundamental g-mode, visible particularly in
Fig.~\ref{f:aabd1_2dslice_mt10g50mm00n03_s-ms}
as a significant brightening between heights of 0 and 20\,km.
In addition, there are several modes with larger horizontal and vertical wave number.
These waves show up together with the first surface granules,
well before the downdrafts ``hit'' the lower boundary.
This indicates that the granular flow as such is responsible for the wave excitation,
and not artifacts related to the way flows at the lower boundary are handled.

Figures~\ref{f:aabd1_2dslice_mt18g50mm00n07_vel_qucallm1orho} and
\ref{f:aabd1_2dslice_mt18g50mm00n07_qucallm1orho}
%\ref{f:aabd1_2dslice_mt10g50mm00n03_vel_qucallm1orho}, and
%\ref{f:aabd1_2dslice_mt10g50mm00n03_qucallm1orho}
demonstrate the location of the dust clouds
and the effect of the thermal inhomogeneities induced by the gravity waves
onto the dust concentration.
The generated small amount of vertical mixing
(the wave motion is mostly reversible)
is sufficient to balance gravitational settling
of dust grains and allow dust clouds to form
in the hotter models.
In addition, dust concentration and cloud thickness are modulated by the waves
because of the induced temperature fluctuations.

Atmospheric gravity waves are a common phenomenon.
On Earth, they are known to form clouds over e.g., the US midwest
plains\footnote{Mesonet, I.~E. 2007, Gravity Wave Movie: \href{http://mesonet.agron.iastate.edu/cool/}{http://mesonet.agron.iastate.edu/cool/}}.
%\citep{IEM2007GravityWaveMovie}.
Their energy release is involved in heating
the exospheres of Jovian planets as observed for Jupiter from Galileo probe results
\citep{Young1998JGR...10322775Y}.

Simulations of convection producing gravity waves in stellar conditions
have a long tradition \citep{Hurlburt1986ApJ...311..563H}.
However, the quantitative estimate of the amplitude
and the true detection of internal gravity waves
can be a difficult task, even for the well-studied
solar case \citep{Belkacem2009A&A...494..191B}.
The detection of gravity modes that probe the solar core
was announced by \cite{Garcia2007Sci...316.1591G}.

The initial phases of simulation indicate that gravity waves are generated near the
top of the convection zone
\citep[see e.g.,][]{Dintrans2005A&A...438..365D}.
The gravity waves are produced by non-stationary downdrafts
``sucking'' at the stable photospheric layers.
In this way, the downdrafts are able to
inject kinetic energy into the photosphere
and to transport some material from there into the deeper convection zone.
However, there are no obvious ``events'' of wave generation as for p-modes
in the sun \citep{Goode1998ApJ...495L..27G, Stein2001ApJ...546..585S}
or in the simulations of gravity waves generated by
an idealized convection zone embedded between stable layers by
\cite{Dintrans2005A&A...438..365D}.

The mixing efficiency of the waves increases rapidly with height -- steeper
than expected from the mere growths in amplitude caused by the increasing non-linearity.
This could be the dynamical updraft mechanism responsible for the upwelling of
N$_2$ and CO gas observed via the enrichment of CO and depletion of CH$_4$ and
NH$_3$ absorption bands in the spectra of T dwarfs
\citep{Saumon2006ApJ...647..552S, Stephens2009ApJ...702..154S,
  Geballe2009ApJ...695..844G}.

%===============================================================================
\subsection{Convection within dust clouds}

The fluctuations in the dust concentration in the 1800\,K model
in Fig.~\ref{f:aabd1_2dslice_mt18g50mm00n07_vel_qucallm1orho}
are mainly induced by up and down motions of gravity waves
that provide an inefficient mixing that balances the settling of dust grains.
However, when the optical thickness of the dust clouds becomes sufficiently high,
convective motions \emph{within the dust clouds} start to develop
and provide more efficient mixing of material
(cf.\ the dust concentration of the 1000\,K model in
Fig.~\ref{f:aabd1_2dslice_mt10g50mm00n03_qucallm1orho}).
However, in the snapshots the fluctuations and flows
due to the waves somewhat obscure the dust cloud convection,
whereas the overturning motions are clearly visible in movies
and have a different signal in a $k$-$\omega$~diagram.

There are different intermittent processes:
occasionally, material from the dust layers is dredged up to
the layers with relatively low dust concentrations above the clouds.
The grains quickly fall back.
But monomers can remain a while, until they condense into dust
at the top of the cloud deck.
The cloud layer thickness varies not only with the wave on a timescale below one minute
but also in irregular cycles on timescales of hours.
The irregularity and amplitude increases with decreasing effective temperature.

During the initial phases of a simulations, a violent thin cloud convection zone
develops for a limited time until the model is relaxed. This phenomenon
relates to differences between our start model and the final outcome.
However, on actual brown dwarfs large-scale flows might cause an imbalance
in the local dust concentration that leads to a similar localized
enhanced cloud activity.

%===============================================================================
\subsection{Dust and stratification}

\begin{figure*}
\centering
\includegraphics[]{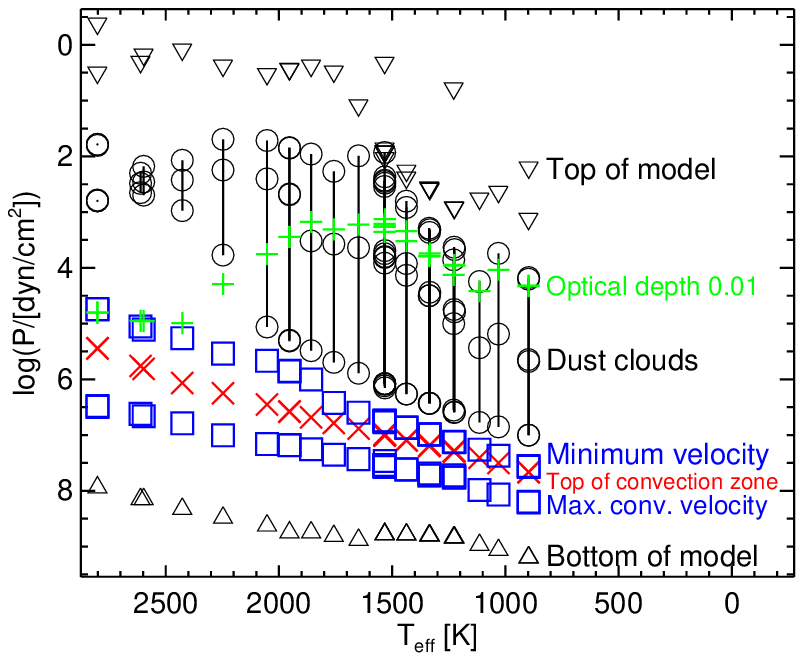}
\includegraphics[]{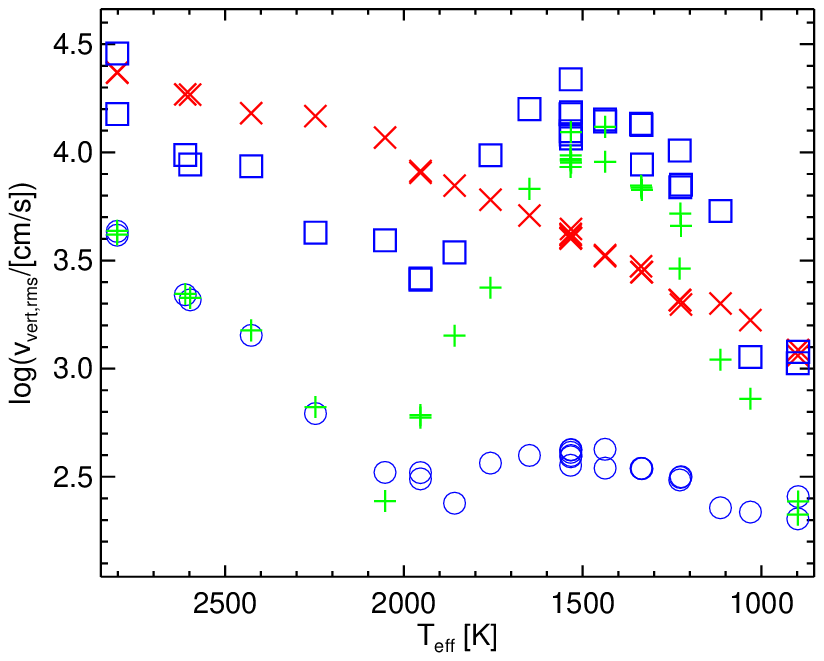}
\includegraphics[]{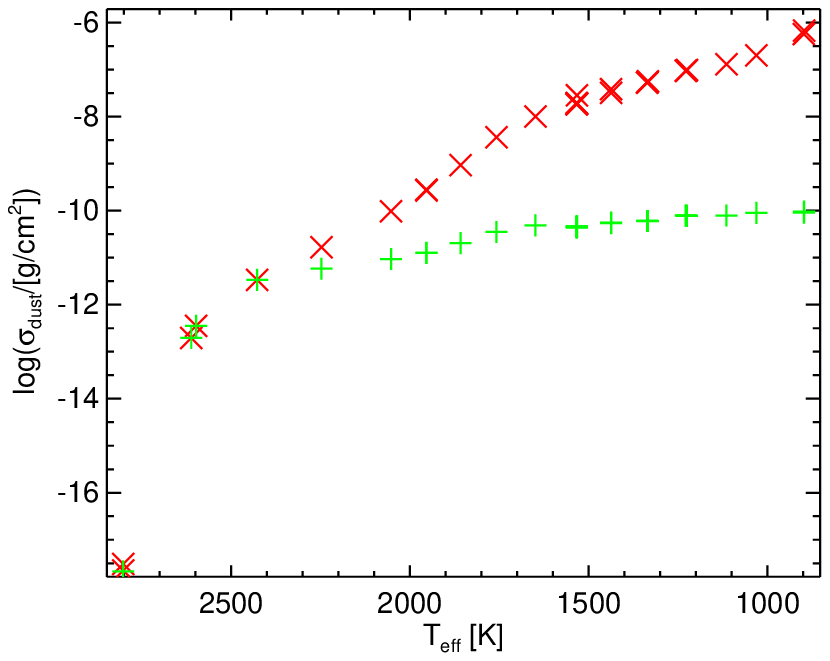}
\includegraphics[]{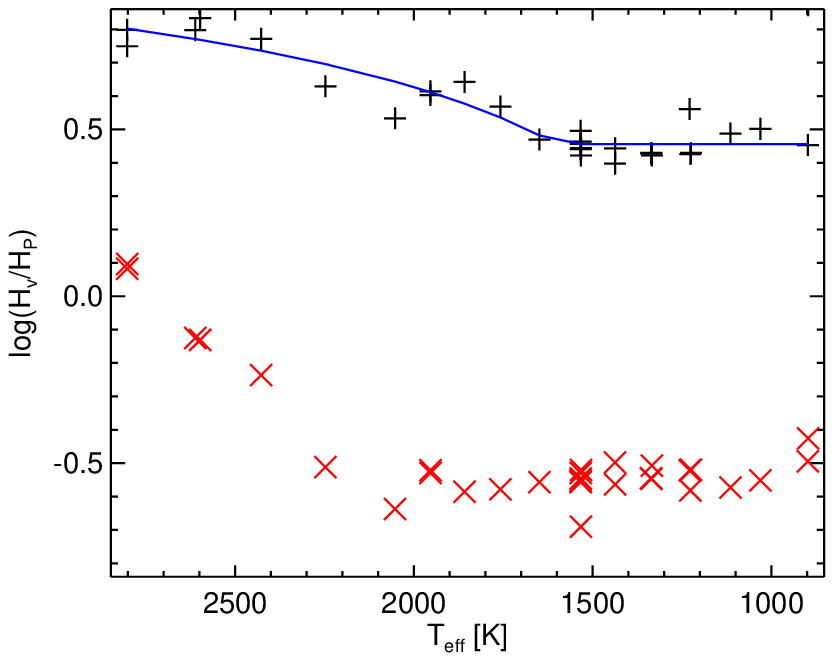}
\caption{
Various quantities plotted versus effective temperature for all models
in Table~\ref{t:modelparam}.
{\bf
Top left;
logarithm of pressure for various points of interest}:
black triangles: top and bottom of each model,
red crosses: top of convectively unstable layers,
lower set of blue squares: point with maximum convective velocity $v_\mathrm{vert,rms}$,
black circles: upper and lower boundary of cloud layers
(region where the dust concentration lies above 10$^{-6}$)
and point of maximum dust concentration,
green plus signs: layers with $\tau_\mathrm{Rosseland}$=$10^{-2}$.
{\bf Top right; rms value of vertical velocity $v_\mathrm{vert,rms}$}:
red crosses: maximum convective velocities,
blue squares: maximum wave velocities,
blue circles: minimum velocity in between,
green plus signs: velocity at $\tau_\mathrm{Rosseland}$=$10^{-2}$.
{\bf Bottom left; total amount of dust}:
red crosses: total amount of dust in model,
green plus signs: dust above layers with $\tau_\mathrm{Rosseland}$=$10^{-2}$.
{\bf Bottom right; scale height of rms of vertical velocity}:
black plus signs: approximate scale height of increase of wave velocities with height,
blue line: fit according to Eq.~(\ref{e:HvoverHp}),
red crosses: scale height of exponentially declining overshoot velocities.
}
\label{f:aabd_seqavgstat_q_Teff}
\end{figure*}

In the top left panel in Fig.\ref{f:aabd_seqavgstat_q_Teff}, we show
the location of the dust clouds (circles connected by vertical lines) relative
to the underlying gas convection zone (located below the crosses).

The mixing processes within the dust cloud layers have
different height regimes.
At the bottom of the dust clouds, the temperature varies
around the condensation value but there is little mixing.
With our dust scheme, which assumes the presence of nuclei everywhere
where dust or monomers are present, dust forms and evaporates during
these temperature cycles
(see the dust concentration at $z$$\sim$10\,km in
Fig.~\ref{f:aabd1_2dslice_mt10g50mm00n03_qucallm1orho}).
The dust formation would be more difficult if new dust grains had
to nucleate, because that would require some level of supersaturation.
Within the clouds, material is mixed by gravity waves and/or convection
(depending on effective temperature).
The top of the clouds is sharp but inhomogeneous due to (sometimes braking) waves
and cloud convection.
Above the cloud layers, there are still mixing flows that try to equalize the
concentration of monomers with height. The concentration value depends
on the efficiency of mixing, dust formation, and dust settling in the cloud layers
below (Fig.~\ref{f:aabd_seqavgstat_q_log10_P}, bottom panels).

Dust clouds have a strong effect on the thermal structure
(Fig.~\ref{f:aabd_seqavgstat_q_log10_P}, top right panel):
there is a fairly shallow temperature slope beneath the cloud layers
with values of about 1600\,K
because of the greenhouse effect,
a rapid drop within the clouds due to the large dust opacities,
that can even drive cloud convection,
low temperatures (with values around 1000\,K and small variations)
in the mostly dust-free upper atmosphere,
and in some cases a small increase at the top of the models
of about 100\,K
because of the dissipation of kinetic wave energy.
At some height above the cloud, gravitational settling of dust grains becomes more
efficient than mixing. The dust density drops rapidly and with it
dust opacity and temperature, causing a rather sharp (but variable in space and time)
upper boundary of the clouds.
The concentration of dust and monomers (material that potentially can form dust)
in Figs.~\ref{f:aabd1_2dslice_mt18g50mm00n07_qucallorho}
and \ref{f:aabd1_2dslice_mt10g50mm00n03_qucallorho}
shows complete mixing in the convection zone,
depleted layers at the top of the atmosphere (due to gravitational settling),
and a partially mixed region in-between.

% ..............................................................................
% ... aabd1_seqavgstat_pos_plot.pro ...
% ... aabd1_seqavgstat_v3_plot.pro ...
% ... aabd1_seqavgstat_totaldust_plot.pro ...
% ... aabd1_seqavgstat_scaleheight_plot.pro ...

\clearpage
%===============================================================================
\subsection{Effective temperature dependency}

We summarize the dependence of our model properties on effective temperature
in Fig.~\ref{f:aabd_seqavgstat_q_Teff}.
The upper right panel shows that the atmospheric velocities (squares) do not
follow the monotonic decrease in the convective velocities (crosses) up to
the lowest effective temperatures, but rise after a minimum around $T_\mathrm{eff}=1900\,K$.
However, the cloud thickness (top left panel) and cloud mass (bottom left panel)
increase monotonically with decreasing effective temperature.
The cloud extension with $T_\mathrm{eff}$ is slightly erratic
because the models have not perfectly converged.
At higher effective temperatures, the thin see-through clouds allow a view of
the upper layers of the convection zone (plus signs in both top panels),
while at the lower-$T_\mathrm{eff}$ end most of the cloud mass sits below
the visible layers.

The rms vertical velocity, which is often used in static model atmospheres to estimate line 
broadening, non-equilibrium chemistry due to mixing, or cloud formation, can be
parametrized  as a function of effective temperature for our sequence of
30 models. Hence, with the logarithmic normalized temperature
$x=\log ( \min(\max(T, 900\,\mathrm{K}), 2800\,\mathrm{K})/1K)$,
we obtain for the velocity scale height of the wave amplitude
(bottom right panel in Fig.~\ref{f:aabd_seqavgstat_q_Teff})
%yslope=2.867 > (-47.82+15.78*x) (old)
%yslope=2.855 > (-43.1+14.34*x
\begin{equation}\label{e:HvoverHp}
  H_v/H_p=\max(2.855, -43.1+14.34\,x) \enspace ,
\end{equation}
and for the logarithmic ratio
of maximum convective velocity
to wave amplitude extrapolated to this layer
%yratio=(519.468 - 528.211*x + 178.636*x^2 -20.1343*x^3) > (-26.516 + 7.4146*x) (old)
%yratio=(835.684-828.156*x+273.426*x^2-30.1148*x^3) > (-25.496+7.1104*x)
\begin{eqnarray}
  \log r_v=\max( \!\!\! & 835.684 - 828.156 x + 273.426 x^2 -30.1148 x^3, & \, \nonumber \\
                        & -25.496 + 7.1104 x  ) \enspace , &
\end{eqnarray}
where the first expression in the max functions is a good fit
for temperatures below approximately 2000\,K,
and the second expression for temperatures above.
To recover a crude estimation of the mixing efficiency based on these formulae,
we propose to compute the amplitude $V_\mathrm{max}$ and
position $P_\mathrm{max}$ of the maximum convective velocity
with e.g.,\ the Mixing-Length Theory,
to add $\log r_v$ to the logarithmic velocity amplitude,
and to extrapolate from this starting point
the wave amplitude with $H_v/H_p$ into the atmosphere,
%logVfit=(vmaxconv2[i]+yratio[i]) - (xp-vmaxconv1[1])/yslope[i]
\begin{equation}\label{eq:VmixfromP}
 \log V = \log V_\mathrm{max} + \log r_v - (\log P - \log P_\mathrm{max}) / (H_v/H_p)  \enspace ,
\end{equation}
where $P$ is the pressure.
We note, that for the models with $T_\mathrm{eff}$=900\,K and 1000\,K
the velocities drop after reaching a maximum of about 10\,m\,s$^{-1}$,
while the rise in the velocities of the hotter models appears to be limited
only by the top of the computational domain.

The rise of velocities above the convection zone
becomes steeper with decreasing temperature
(bottom right panel in Fig.~\ref{f:aabd_seqavgstat_q_Teff}).
But the amplitudes in the atmosphere
(Figs.~\ref{f:aabd_seqavgstat_q_log10_P}
and \ref{f:aabd_seqavgstat_q_Teff}, top right)
fall from models with high effective
temperatures to models with $T_\mathrm{eff} \approx 2000$\,K.

For even cooler models, the amplitudes increase again.
This increase is not because of the velocities in the underlying gas convection zone
that decline steadily as effective temperature decreases.
Instead, below 2000\,K, the clouds have grown
to such a large vertical thickness and density
that cloud convection begins -- with effects onto the atmospheric velocities
and temperature structure that increase with further decreasing effective temperature.
We attribute the rise in velocity for low-temperature models mainly to
the emergence and growth of cloud convection.

Therefore, the dust clouds also affect the waves:
when they become extended enough they split the atmospheric cavity for
gravity waves into two separate zones by
generating an entropy plateau within the atmosphere
(Fig.~\ref{f:aabd_seqavgstat_q_log10_P}).
At $T_\mathrm{eff}$ below about 1000\,K, the gravity waves are
trapped mainly inside the region between the two convection zones.
The layers above the dust convection zone exhibit only small wave amplitudes.

The convective-radiative boundary becomes steeper, hence harder,
with lower effective temperature, easing the gravity wave generation.
In addition, there is a slight change in the topology of granules:
in the hotter models, the downdrafts that delimit the granules are of roughly similar
strength and merge occasionally,
while in the cooler models just a few (2 or 3) ``super downdrafts'' dominate and
absorb the smaller ones that form on top of the granules.
This process occurs with higher frequency than would be expected if the
merging type were the same as in the hot models --
with possible consequences for the interaction between convection and waves.

% --- Mail from 2009-02-10 16:17 ---
%
%A (relatively good) linear fit to logV_expon over the entire range
%IDL> print,cslope  ; from top to bottom: a0, a1
%     -0.47169217
%      0.72636354
%x=log(Teff/1000)
%y=a0+a1*x
%

%A (relatively good) 2nd order fit to logV_diff over the range 900K - 2000K
%IDL> print,cratio
%     -0.95575941
%     -0.62276429
%      -7.4471689
%x=log(Teff/1000)
%ylow=a0+a1*x+a2*x^2

%A (relatively good) 2nd order fit to logV_diff over the range 2000K - 2800K
%IDL> print,cratio
%      -7.7574785
%       26.100723
%      -24.679659
%x=log(Teff/1000)
%yhigh=a0+a1*x+a2*x^2
%y=max(ylow, yhigh)

%#########################################################################################
\section{Discussion \label{s:Discussion}}

%===============================================================================
\subsection{Parameter dependence}

The details of the wave generation process
and the type, amplitude, interaction, long-term evolution, and spectrum
of the waves are complex and may have connections to numerical settings.
Some small-amplitude pressure waves in the hottest models are emitted from the
non-stationary downdrafts as expected and become
invisible at intermediate temperatures.
However, in the lowest-temperature models ($T_\mathrm{eff}$$\le$1300\,K)
the signature of high-frequency p-modes unexpectedly showed up in the pressure fluctuations,
where g-modes and convection have only a small signal due to their nearly incompressible
nature and small amplitude.
The p-modes contribute to the velocities only close to the very top of the computational box.
Their amplitude depends sensitively on viscosity and the position of the top boundary.
They vanish, when the Courant number is reduced from 0.4 to about 0.3, depending on
other model details
(the usual stability criterion sets an upper limit at 0.5 for the 2D models).
The models that we used for our final analysis show no or only traces of these p-modes.

An early version of the models showed (in addition to the ``normal'' spectrum of gravity
waves that occur as soon as convection sets in)
after the simulation had run for a long time a slowly exponentially growing gravity wave
in the fundamental mode.
It grew until the code crashed because of too steep velocity gradients at the top of the box.
It had relatively little effect on mixing, but induced temperature fluctuations
modulating the dust concentration.
Limiting the model depth, and using both a finer vertical grid and a smaller Courant number
prevented an exponential growth of the mode.
However, the mode itself is still present and quite prominent in the cooler models.

% ..............................................................................
% seqavgstat_plot.pro
% aabd1_seqavgstat_log10_v3rms_log10_P_plot.pro
\begin{figure}
\centering
\includegraphics[]{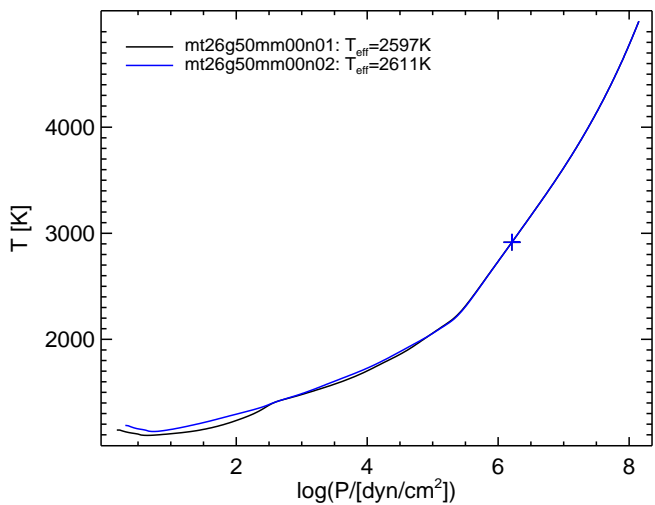}
\caption{Mean temperature versus logarithm of pressure for two models
with $T_\mathrm{eff}$$\approx$2600\,K, $\log g$=5, but different opacity tables.
}
\label{f:aabd_seqavgstat_mt26_T_log10_P}
\end{figure}

% ..............................................................................
% aabd1_seqavgstat_log10_v3rms_log10_P_plot.pro
\begin{figure}
\centering
\includegraphics[]{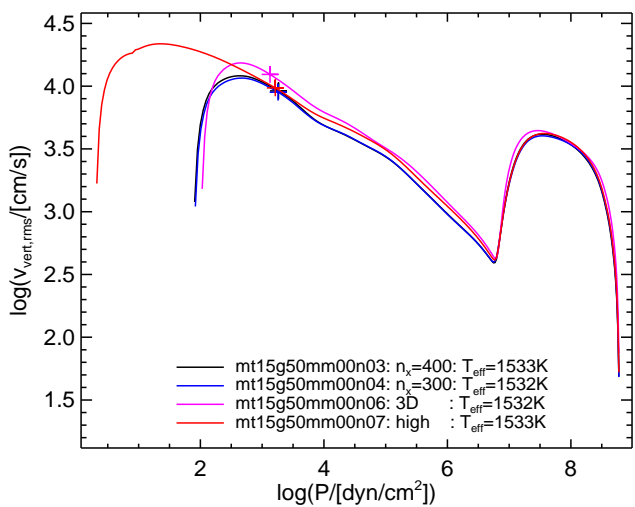}
\caption{The rms value of vertical velocity versus logarithm of pressure for models
with $T_\mathrm{eff}$$\approx$1500\,K, $\log g$=5, but differences in the geometry
(horizontal grid points, vertical extent, dimension).
}
\label{f:aabd_seqavgstat_mt15_log10_v3rms_log10_P}
\end{figure}

We varied several numerical parameters to check their influence.
By considering a pair of models with $T_\mathrm{eff}$=1300\,K, one in single,
the other in double precision, we found hardly any difference at all in the
resulting mean properties (velocities and temperature).
However, at $T_\mathrm{eff}$$\le$1100\,K the density fluctuations in
the convection zone become so small that they cannot be resolved
properly using single precision: after a transient phase, convection dies
out leaving only small-scale low-amplitude velocity fluctuations in
the ``convection zone'' that are due to round-off errors.  Therefore, all runs
with $T_\mathrm{eff}$$\le$1100\,K were performed in double precision from
the initial tests on.

For a 1500\,K model, we decreased our standard horizontal resolution
by going from 400 to 300 horizontal grid points and found no
noticeable difference in the mean structures, although the thin
convective downdrafts and some small-scale cloud structures are
somewhat less well resolved.

For the generation of our binned opacity tables, the bins were optimized 
for three reference atmosphere cases ($T_\mathrm{eff}$=1000K, 1800K, 
and 2800K with log\,$g$=5.0 and solar metallicity). The resulting tables 
are most accurate for model parameters close to these values.
In Fig.~\ref{f:aabd_seqavgstat_mt26_T_log10_P}, we show the temperature
structure of two models with $T_\mathrm{eff}$=2600\,K using an opacity table
made with a 1800\,K reference atmosphere in mt26g50mm00n01,
and with a more appropriate 2800\,K reference atmosphere in mt26g50mm00n02.
The resulting differences are most important for the outermost layers, but remain 
negligible for determining the mixing mechanisms in these atmospheres.

Figure~\ref{f:aabd_seqavgstat_mt15_log10_v3rms_log10_P} shows the rms
velocity for four 1500\,K runs with different positions of the upper
boundary.
As expected, the more extended model has a larger peak velocity, while
the agreement between the two curves is good in the lower atmospheric
layers and excellent in the convection zone.
We tried the same experiment at $T_\mathrm{eff}$=900\,K, 1200\,K and, 2800\,K
with similar agreement between the pairs of curves.

Simulations that are not yet complete and that will be presented in
future publications include a sequence with other gravity values that
shows no qualitative change in the outcome, although convective
velocities, wave amplitudes, and dust-formation rate equations
noticeably depend on gravity: the dependence of the flow field and
the cloud thickness on effective temperature will be different
at other gravities.

A 3D model with $T_\mathrm{eff}$=1500\,K (after taking about half a year
to cover 4~hours of stellar time with six CPUs)
has completed the transition from the initial 2D configuration to
a fully 3D flow pattern.
We see only small changes in the wave amplitudes and convective velocities
relative to a 2D run (Fig.~\ref{f:aabd_seqavgstat_mt15_log10_v3rms_log10_P}).
The 3D model data are in close agreement with the 2D models in the scatter plots
in Fig~\ref{f:aabd_seqavgstat_q_Teff}.
Only its overshoot scale height (bottom right panel, $H_v$$=$$0.2\,H_p$)
is noticeably smaller than the average of the 2D models
at $T_\mathrm{eff}$=1500\,K ($H_v$$=$$0.3\,H_p$).

The exploration of other parameters such as grain size and other types
of dust awaits further simulations, using a more detailed cloud model
(multi-size-bin scheme, in preparation).

%===============================================================================
\subsection{Comparison with previous simulations}

\cite{Ludwig2002A&A...395...99L, Ludwig2006A&A...459..599L} studied the
structure of mid M-type, dust-free atmospheres including the mixing properties
of macroscopic flows with 3D hydrodynamical simulations. Our
$T_\mathrm{eff}$$\approx$$2800\,\mathrm{K}$ model coincides in temperature with the
coolest model of Ludwig et al., and this allows a comparison with the results.
While the maximum rms vertical velocities in the convectively unstable layers
turn out to be similar,
the atmospheric-wave-dominated velocities are about 50\,\% higher in Ludwig et al.
Importantly, the scale height of the decline in the convective
velocity field amplitude into the stably stratified layers is found to be similar
($H_v/H_p = 1.2$ this work, $1.1$ in the work of Ludwig et al.). At first
sight, this may appear surprising considering the systematic differences
expected between simulations conducted in 2D and 3D as discussed by
\citet{LudwigNordlund2000}: in 2D the efficiency of wave generation is usually
higher, and the transition between convectively stable and unstable regions is
more gradual. However, the work of Ludwig \&\ Nordlund refers to higher
Mach-number flows, and pressure waves, not gravity waves, which are relevant
here. In the higher Mach-number conditions studied by Ludwig \&\ Nordlund,
towards lower Mach-numbers the sharpness
in the stable-unstable transition becomes more similar in the 2D and 3D
simulations, so that the similarity in the M dwarf regime appears plausible.
This is also borne out by a comparison with a 3D model compiled for
2800\,K, which has a very similar velocity profile to
its 2D counterpart.

Ludwig and collaborators had reasons to believe that the gravity waves present
in their models were an artifact of the lower boundary condition. Moreover,
they argued that the mixing efficiency is too small -- because of both low (i.e.,
linear) amplitude and the insufficient shearing -- to produce small-scale turbulence
due to Kelvin-Helmholtz instabilities. Since in their models, convective overshoot
was potentially able -- when extrapolated to lower $T_\mathrm{eff}$ -- to keep dust
grains in the optically thin layers, they took assumed that waves are not
important and convective overshooting is sufficient to explain the presence of
dust clouds in brown dwarfs.

The present calculations cover the actual parameter regime of dust harbouring
atmospheres. They show that -- in contrast to expectations motivated by hotter
models -- convective overshoot alone is not capable of keeping dust grains in the
atmosphere. Overshooting motions decline more rapidly towards lower
effective temperatures (Fig.~\ref{f:aabd_seqavgstat_q_Teff}). Our thorough
investigation of the influence of the boundary conditions on the excitation of
the gravity waves indicate that they are indeed intrinsic to the flow
evolution proper and not a numerical artifact. The gravity waves's ability to
mix is indeed low but the waves remain in our hotter local models nevertheless the most
efficient process, accompanied by dust convection in the case of heavy dust
formation. All in all, we consider our present results consistent with the
findings of \cite{Ludwig2002A&A...395...99L, Ludwig2006A&A...459..599L},
but reassign the importance to the mixing by waves.

A complementary approach to ours is pursued by \cite{Helling2004A&A...423..657H}:
rather than on macroscopic scales (pressure scale heights, depth of the atmosphere,
granular diameter), they concentrate on mesoscopic scales.
Their 2D model is about as large (500$\times$500\,m$^2$) as one of our grid cells.
They investigate the influence of driven turbulence,
represented by a set of imposed pressure waves,
on the formation of dust,
particularly in regions where the temperature is slightly too high (T$=$2100\,K)
to allow nucleation in an undisturbed atmosphere.
We agree with their findings that fluctuations in the thermodynamic quantities
can have an influence on the dust formation process.
However, we identify gravity waves and not pressure waves
as important contributors to the mixing in BD and M dwarfs,
in addition to with convection within thick clouds and convective overshoot very
close to the underlying gas convection zone.
An important parameter in their simulations is the Mach number of the induced acoustic waves,
for which they assume values of about 0.1 in 1D models and 1 in 2D models.
However, peak convective Mach numbers
(taking vertical and horizontal velocities into account)
in our models are between 0.1 and 0.01,
and rapidly decrease in the overshoot regions where high-temperature dust might form.
The amplitude of turbulent structures on the grid cell scale and below
-- that we obviously cannot resolve in our models --
would be even smaller.
And only a tiny fraction of the energy can be expected to be transformed into
pressure waves under these nearly incompressible low-Mach-number conditions.
Therefore, based on our simulations we cannot justify the assumption
of almost sonic pressure waves in the atmospheres of brown dwarfs
as made in \cite{Helling2004A&A...423..657H}.

%===============================================================================
\subsection{Diffusion coefficient estimate}

% ..............................................................................
% aabd1_seqavgdiffusion_plot.pro
\begin{figure}
\centering
\includegraphics[]{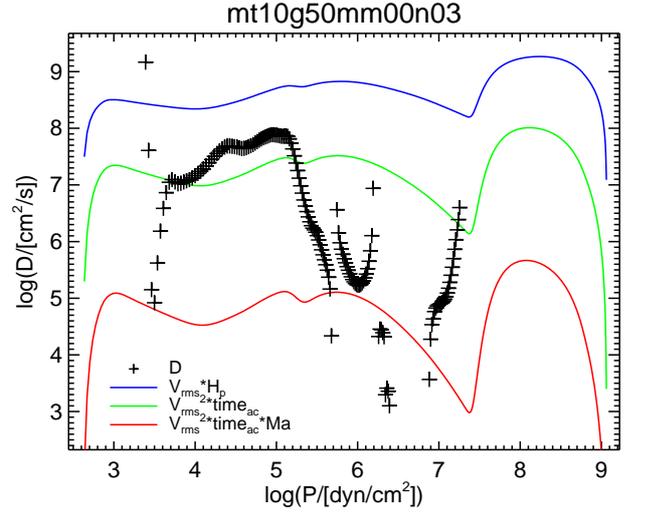}
\caption{Logarithm of diffusion coefficient versus logarithm of pressure
for a model with  $T_\mathrm{eff}$=1000\,K according to
Eqs.~(\ref{e:DiffVHp}) to (\ref{e:DiffMaV2t}).
The black crossed indicated estimates derived from the the averaged vertical
fluxes and concentrations of dust and monomers.
}
\label{f:aabd_seqavgdiffusion_log10_D_log10_P}
\end{figure}

% ..............................................................................
% ... aabd1_seqavgstat_log10_v3rms_log10_P_plot.pro ...
\begin{figure}
\centering
\includegraphics[]{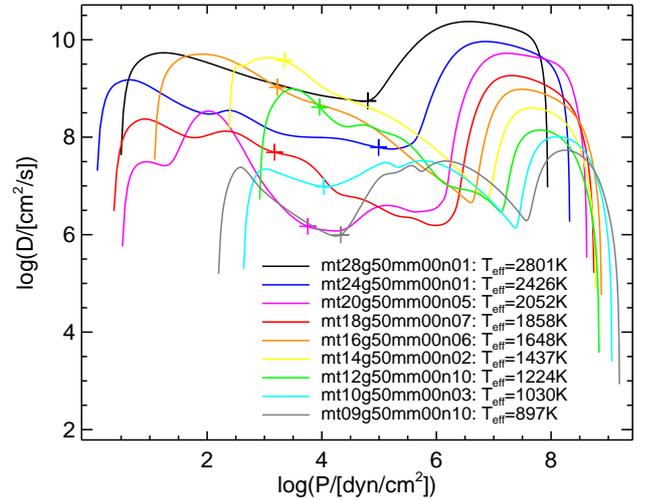}
\caption{Logarithm of diffusion coefficient $D$ (or $K_{zz}$)
according to Eq.~(\ref{e:DiffV2t})
versus logarithm of pressure for the same set of models as in Fig.~\ref{f:aabd_seqavgstat_q_log10_P}.
}
\label{f:aabd_seqavgstat_D_log10_P}
\end{figure}

One can model the mixing of material by macroscopic flows -- on average --
as a diffusion process.
However, in the hydrodynamical models there is a
correlation between the sign of the vertical motions (upward of downward) and
the grain growth.
At the same velocity amplitude, the mixing efficiency of convective overturning flows
is also much higher than that of (nearly reversible) wave motions,
causing errors in the translation from the rms velocities to the actual mixing efficiency.
The mixing efficiency can however be estimated from the rms vertical velocity
of our model sequence as in Eq.~(\ref{eq:VmixfromP}).
And the diffusion coefficient can be estimated 
from the local vertical velocity and the pressure scale height $H_p$ as typical length scale via
\begin{equation}\label{e:DiffVHp}
  D \propto V \, H_p \enspace .
\end{equation}
However, the waves have a varying amplitude with height
and therefore the typical length scale is not constant.
On the other hand, their period is rather close to the Brunt-V{\"a}is{\"a}l{\"a} period,
which can be used as a characteristic timescale.
Using the similar acoustic period, $t_\mathrm{ac}=2H_p/c_\mathrm{sound}$, we obtain
\begin{equation}\label{e:DiffV2t}
  D \propto V^2 \, t_\mathrm{ac} \propto \mathrm{Ma} \, V \, H_p \enspace ,
\end{equation}
where the Mach number $\mathrm{Ma}=V/c_\mathrm{sound}$.
To take into account the increase in mixing with increase in non-linearity, one could
multiply with the Mach number again to obtain
\begin{equation}\label{e:DiffMaV2t}
  D \propto \mathrm{Ma}^2 \, V \, H_p \enspace .
\end{equation}
Profiles for the diffusion coefficients according to the
Eqs.~(\ref{e:DiffVHp}) --- (\ref{e:DiffMaV2t})
(replacing ``$\propto$'' by ``$=$'')
are displayed in Fig.~\ref{f:aabd_seqavgdiffusion_log10_D_log10_P}.
Additional crosses mark estimates based on the horizontal and temporal averages
of vertical flux and density profiles of dust plus monomers.
Because the flux is divided by the vertical derivative of the concentration,
which can be very small, these values are not well-behaved everywhere.

The diffusion coefficient in brown dwarfs is not a ``constant of nature''
but depends on the physical process driving the mixing, the height in the atmosphere,
and the effective temperature.
A lower gravity would lead
(via increased convective velocities and larger time and spatial scales)
to high diffusion coefficients.

Scaling relations such as Eqs.~(\ref{e:DiffV2t}) and (\ref{e:DiffMaV2t}) can
serve as a first step in describing the diffusion properties in ultracool
atmospheres in greater detail. In addition, they can easily be
translated into classical atmosphere codes. 
In an earlier study, we have implemented a non-equilibrium chemistry 
model in the \texttt{PHOENIX} BT-Settl code, using the diffusion
coefficients derived from the overshoot contribution of the RHD
simulations only. In atmosphere models of the T1 brown
dwarf $\epsilon$~Indi~Ba based on this approach, we obtain characteristic
values of 10$^7 \le D \le 10^8$ cm$^2$\,s$^{-1}$ for the transition
region from CO- to CH$_4$-dominated carbon chemistry.  
We found the resulting non-equilibrium abundances of CO in the
line-forming region, using an updated and more efficient reaction
model than \citet{Saumon2006ApJ...647..552S}, to just slightly
underestimate the observed CO line strengths in this benchmark
transition dwarf \citep{King2009arXiv0911.3143K}. 
This indicates that somewhat more efficient mixing than provided by
overshoot alone is required, thus supporting an additional
contribution from gravity waves.
The latest revision of the BT-Settl models beeing
tested to include the effect of both dust mixing and
CE departures, also find that Eq.~(\ref{e:DiffV2t}) provides a close match
to observational constraints. 

In a more phenomenological approach \citet{Saumon2006ApJ...647..552S}
and \citet{Stephens2009ApJ...702..154S} explored the effects of a 
constant eddy diffusion coefficient above the Schwarzschild boundary
on observable departures from nitrogen and carbon equilibrium
chemistry. They found best-fit values ranging from 10$^2$ to 10$^6$
cm$^2$\,s$^{-1}$, although based on slower reaction rates (see above). 
Allowing for the differences in the reaction scheme, their results
thus agree with the range of values that we find for the diffusion
coefficient (crosses in
Fig.~\ref{f:aabd_seqavgdiffusion_log10_D_log10_P}) 
in the region spanning the top of the overshoot layers,
the gravity-wave region,
and the base of the cloud layers ($7.2 > \log P > 5.5$).
Within the clouds, we find larger values.

%===============================================================================
\subsection{Brightness variations}

% ..............................................................................
% ... aabd1_seqavgstat_intrms_plot.pro ...
\begin{figure}
\centering
\includegraphics[]{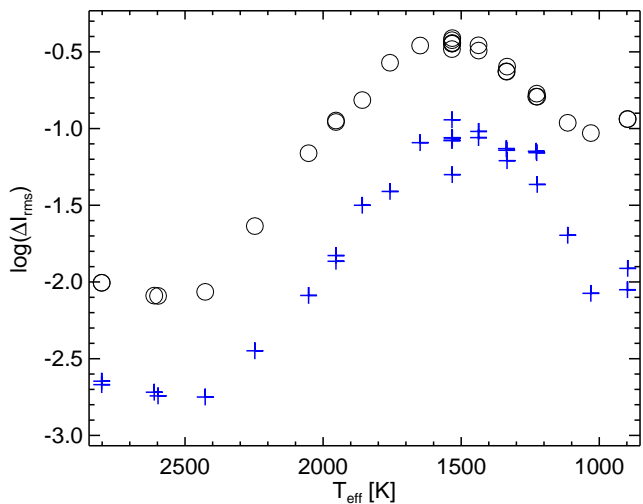}
\caption{
Logarithm of relative bolometric flux intensity contrast plotted versus effective temperature for all models
in Table~\ref{t:modelparam}.
{\bf Top curve (circles)}:
total contrast (spatial plus temporal contribution).
{\bf Bottom curve (plus signs)}:
temporal variations only.
}
\label{f:aabd_seqavgstat_intrms_Teff}
\end{figure}

Our simulations show temporal intensity variations for a wide range
of timescales from half a minute to hours.
We find low-amplitude (less than 1\%, see Fig.~\ref{f:aabd_seqavgstat_intrms_Teff})
and short-period variability (1\,min)
due to the gravity waves producing temperature fluctuations
and modulating the dust density and the vertical thickness of the clouds.
Relatively short-lived (several minute) phenomena are the occasional
dredge-up or outburst of material above the clouds, where dust
falls back rapidly while monomers remain much longer.
In the coolest models, the gravity waves are less visible in the
intensity fluctuations, which are dominated instead by
aperiodic variations on the scale of hours
with an amplitude of a few per cent.
But these results barely reach the scales of
the observed variability of L-type brown dwarfs, which is often
aperiodic (scales of hours to days) and of low amplitude
(mmags)
as reported by \cite{Gelino2002ApJ...577..433G}.
Still, the spatial intensity contrast found in the models
(Fig.~\ref{f:aabd_seqavgstat_intrms_Teff})
with significant dust layers is significantly higher than the
contrast that would be induced by granulation alone
(the small contrast for models above about 2500\,K).

Our results are only those
for a patch of the surface and the variability amplitudes obtained will
average over the rest of the surface. To determine the brown dwarf
surface distribution of clouds, one must go beyond the
present simulations to 3D models that are as large as possible and include
rotation effects.

We have neglected the effects of rotation despite the rapid rotational
periods of brown dwarfs (P $\le 4$\, hrs). The convective turnover
time in the box is several minutes, which is short in comparison.
Neither the surface granules, nor our rms velocities, should
therefore be severely affected.  But granules could move with a
global meridional flow.  Other global flows caused by rapid rotation may
exist that could move the dust around.  

A cloud cover disruption has indeed
been suggested as a possible additional cause -- the
cloud layers sinking relative to the line forming layers -- of the L-T
spectral transition \citep{AckermanMarley2001ApJ...556..872A} that
could lead to weather phenomena and spectroscopic variability.

%#########################################################################################
\section{Conclusions \label{s:Conclusions}}

We have performed radiation hydrodynamics simulations with CO5BOLD
of a sequence of brown dwarf atmospheres
extending previous studies to lower temperatures.
The numerical model includes
a simple treatment of the formation and destruction of dust,
as well as its gravitational settling and advection,
and also the interaction with the radiation field.

We provide a fit to the rms velocity in the atmosphere
that can be used to estimate the mixing.
The convective velocities fall significantly from the peak value
inside the convection zone
to the top of the unstable layers, and even further into the overshooting region.
However, the scale height of
exponentially decreasing overshoot velocities
is so small that they do not induce
significant mixing in the cloud layers.
Above a local minimum in the vertical velocities,
gravity waves dominate in the hotter models
with an amplitude and mixing efficiency that increase rapidly with height,
enough to balance the gravitational settling of dust.
The wave amplitude decreases with decreasing effective temperature.
In the cooler models, the dust layers are thick enough to cause
convection within the clouds leading to efficient mixing within the cloud layers.

Models with high effective temperatures
($2500\,\mathrm{K} < T_{\rm eff} < 2800\,$K)
show a high-altitude haze of optically thin
forsterite clouds. At lower effective temperatures ($T_{\rm eff} <
1400\,$K), thick and dense forsterite clouds exist but mostly below
the visible layers, which are essentially depleted of the material that
went into the dust. For intermediate effective temperatures, dust is
an important opacity source in the atmosphere. This agrees
with observations by \cite{Golimowski2004AJ....127.3516G}, which place
i) the onset of important refractory element depletion, where both TiO and VO
bands weaken in spectra because of condensation of titanium and vanadium,
and greenhouse effects at about $2500\,$K,
ii) the maximum greenhouse effects at about $1800\,$K (M to L transition), and
iii) the transition between dust-rich and dust-free brown dwarfs (L to T
transition) at around $1450\,$K.
We therefore feel confident that the
mixing efficiency determined by our simulations is adequate.
Although an investigation of the spectral properties of the models
exceeds the scope of this paper, the formulae that we provide
for the velocity field will allow the discrimination between diverse cloud
model assumptions for brown dwarfs and planetary atmospheres.

%\clearpage
%###############################################################################
\begin{acknowledgements}
We acknowledge financial support from
the {\sl Agence Nationale de la Recherche} (ANR), and the
{\sl ``Programme National de Physique Stellaire''} (PNPS) of CNRS/INSU, France.
The computations were performed at the
{\sl P\^ole Scientifique de Mod\'elisation Num\'erique} (PSMN) at the
{\sl \'Ecole Normale Sup\'erieure} (ENS) in Lyon.
\end{acknowledgements}

%###############################################################################
\bibliographystyle{aa}    % style aa.bst
\bibliography{aa_redsg}

\begin{thebibliography}{52}
\expandafter\ifx\csname natexlab\endcsname\relax\def\natexlab#1{#1}\fi

\bibitem[{{Ackerman} \& {Marley}(2001)}]{AckermanMarley2001ApJ...556..872A}
{Ackerman}, A.~S. \& {Marley}, M.~S. 2001, \apj, 556, 872

\bibitem[{{Alexander} {et~al.}(1997){Alexander}, {Allard}, {Tamanai}, \&
  {Hauschildt}}]{Alexander1997Ap&SS.251..171A}
{Alexander}, D.~R., {Allard}, F., {Tamanai}, A., \& {Hauschildt}, P.~H. 1997,
  \apss, 251, 171

\bibitem[{{Allard} {et~al.}(2001){Allard}, {Hauschildt}, {Alexander},
  {Tamanai}, \& {Schweitzer}}]{Allard2001ApJ...556..357A}
{Allard}, F., {Hauschildt}, P.~H., {Alexander}, D.~R., {Tamanai}, A., \&
  {Schweitzer}, A. 2001, \apj, 556, 357

\bibitem[{{Asplund} {et~al.}(2000){Asplund}, {Ludwig}, {Nordlund}, \&
  {Stein}}]{Asplund2000A&A...359..669A}
{Asplund}, M., {Ludwig}, H.-G., {Nordlund}, {\AA}., \& {Stein}, R.~F. 2000,
  \aap, 359, 669

\bibitem[{{Baraffe} {et~al.}(1995){Baraffe}, {Chabrier}, {Allard}, \&
  {Hauschildt}}]{Baraffe1995ApJ...446L..35B}
{Baraffe}, I., {Chabrier}, G., {Allard}, F., \& {Hauschildt}, P.~H. 1995,
  \apjl, 446, L35+

\bibitem[{{Belkacem} {et~al.}(2009){Belkacem}, {Samadi}, {Goupil}, {Dupret},
  {Brun}, \& {Baudin}}]{Belkacem2009A&A...494..191B}
{Belkacem}, K., {Samadi}, R., {Goupil}, M.~J., {et~al.} 2009, \aap, 494, 191

\bibitem[{{B{\"o}hm-Vitense}(1958)}]{BohmVitense1958ZA.....46..108B}
{B{\"o}hm-Vitense}, E. 1958, Zeitschrift fur Astrophysik, 46, 108

\bibitem[{{Burrows} {et~al.}(2006){Burrows}, {Sudarsky}, \&
  {Hubeny}}]{Burrows2006ApJ...640.1063B}
{Burrows}, A., {Sudarsky}, D., \& {Hubeny}, I. 2006, \apj, 640, 1063

\bibitem[{{Cushing} {et~al.}(2008){Cushing}, {Marley}, {Saumon}, {Kelly},
  {Vacca}, {Rayner}, {Freedman}, {Lodders}, \&
  {Roellig}}]{Cushing2008ApJ...678.1372C}
{Cushing}, M.~C., {Marley}, M.~S., {Saumon}, D., {et~al.} 2008, \apj, 678, 1372

\bibitem[{{Dintrans} {et~al.}(2005){Dintrans}, {Brandenburg}, {Nordlund}, \&
  {Stein}}]{Dintrans2005A&A...438..365D}
{Dintrans}, B., {Brandenburg}, A., {Nordlund}, {\AA}., \& {Stein}, R.~F. 2005,
  \aap, 438, 365

\bibitem[{{Ferguson} {et~al.}(2005){Ferguson}, {Alexander}, {Allard}, {Barman},
  {Bodnarik}, {Hauschildt}, {Heffner-Wong}, \&
  {Tamanai}}]{Ferguson2005ApJ...623..585F}
{Ferguson}, J.~W., {Alexander}, D.~R., {Allard}, F., {et~al.} 2005, \apj, 623,
  585

\bibitem[{{Fortney} {et~al.}(2006){Fortney}, {Cooper}, {Showman}, {Marley}, \&
  {Freedman}}]{Fortney2006ApJ...652..746F}
{Fortney}, J.~J., {Cooper}, C.~S., {Showman}, A.~P., {Marley}, M.~S., \&
  {Freedman}, R.~S. 2006, \apj, 652, 746

\bibitem[{{Freytag} \& {H{\"o}fner}(2008)}]{Freytag2008A&A...483..571F}
{Freytag}, B. \& {H{\"o}fner}, S. 2008, \aap, 483, 571

\bibitem[{{Freytag} {et~al.}(1996){Freytag}, {Ludwig}, \&
  {Steffen}}]{Freytag1996A&A...313..497F}
{Freytag}, B., {Ludwig}, H.-G., \& {Steffen}, M. 1996, \aap, 313, 497

\bibitem[{{Freytag} {et~al.}(2002){Freytag}, {Steffen}, \&
  {Dorch}}]{Freytag2002AN....323..213F}
{Freytag}, B., {Steffen}, M., \& {Dorch}, B. 2002, Astronomische Nachrichten,
  323, 213

\bibitem[{{Gadun} {et~al.}(2000){Gadun}, {Hanslmeier}, {Pikalov}, {Ploner},
  {Puschmann}, \& {Solanki}}]{Gadun2000A&AS..146..267G}
{Gadun}, A.~S., {Hanslmeier}, A., {Pikalov}, K.~N., {et~al.} 2000, \aaps, 146,
  267

\bibitem[{{Garc{\'{\i}}a} {et~al.}(2007){Garc{\'{\i}}a}, {Turck-Chi{\`e}ze},
  {Jim{\'e}nez-Reyes}, {Ballot}, {Pall{\'e}}, {Eff-Darwich}, {Mathur}, \&
  {Provost}}]{Garcia2007Sci...316.1591G}
{Garc{\'{\i}}a}, R.~A., {Turck-Chi{\`e}ze}, S., {Jim{\'e}nez-Reyes}, S.~J.,
  {et~al.} 2007, Science, 316, 1591

\bibitem[{{Geballe} {et~al.}(2009){Geballe}, {Saumon}, {Golimowski}, {Leggett},
  {Marley}, \& {Noll}}]{Geballe2009ApJ...695..844G}
{Geballe}, T.~R., {Saumon}, D., {Golimowski}, D.~A., {et~al.} 2009, \apj, 695,
  844

\bibitem[{{Gelino} {et~al.}(2002){Gelino}, {Marley}, {Holtzman}, {Ackerman}, \&
  {Lodders}}]{Gelino2002ApJ...577..433G}
{Gelino}, C.~R., {Marley}, M.~S., {Holtzman}, J.~A., {Ackerman}, A.~S., \&
  {Lodders}, K. 2002, \apj, 577, 433

\bibitem[{{Golimowski} {et~al.}(2004){Golimowski}, {Leggett}, {Marley}, {Fan},
  {Geballe}, {Knapp}, {Vrba}, {Henden}, {Luginbuhl}, {Guetter}, {Munn},
  {Canzian}, {Zheng}, {Tsvetanov}, {Chiu}, {Glazebrook}, {Hoversten},
  {Schneider}, \& {Brinkmann}}]{Golimowski2004AJ....127.3516G}
{Golimowski}, D.~A., {Leggett}, S.~K., {Marley}, M.~S., {et~al.} 2004, \aj,
  127, 3516

\bibitem[{{Goode} {et~al.}(1998){Goode}, {Strous}, {Rimmele}, \&
  {Stebbins}}]{Goode1998ApJ...495L..27G}
{Goode}, P.~R., {Strous}, L.~H., {Rimmele}, T.~R., \& {Stebbins}, R.~T. 1998,
  \apjl, 495, L27+

\bibitem[{{Hauschildt} {et~al.}(1997){Hauschildt}, {Baron}, \&
  {Allard}}]{Hauschildt1997ApJ...483..390H}
{Hauschildt}, P.~H., {Baron}, E., \& {Allard}, F. 1997, \apj, 483, 390

\bibitem[{{Helling} {et~al.}(2008{\natexlab{a}}){Helling}, {Ackerman},
  {Allard}, {Dehn}, {Hauschildt}, {Homeier}, {Lodders}, {Marley}, {Rietmeijer},
  {Tsuji}, \& {Woitke}}]{Helling2008MNRAS.tmp.1310H}
{Helling}, C., {Ackerman}, A., {Allard}, F., {et~al.} 2008{\natexlab{a}},
  \mnras, 1310

\bibitem[{{Helling} {et~al.}(2008{\natexlab{b}}){Helling}, {Ackerman},
  {Allard}, {Dehn}, {Hauschildt}, {Homeier}, {Lodders}, {Marley}, {Rietmeijer},
  {Tsuji}, \& {Woitke}}]{Helling2008MNRAS.391.1854H}
{Helling}, C., {Ackerman}, A., {Allard}, F., {et~al.} 2008{\natexlab{b}},
  \mnras, 391, 1854

\bibitem[{{Helling} {et~al.}(2004){Helling}, {Klein}, {Woitke}, {Nowak}, \&
  {Sedlmayr}}]{Helling2004A&A...423..657H}
{Helling}, C., {Klein}, R., {Woitke}, P., {Nowak}, U., \& {Sedlmayr}, E. 2004,
  \aap, 423, 657

\bibitem[{{Helling} {et~al.}(2001){Helling}, {Oevermann}, {L{\"u}ttke},
  {Klein}, \& {Sedlmayr}}]{Helling2001A&A...376..194H}
{Helling}, C., {Oevermann}, M., {L{\"u}ttke}, M.~J.~H., {Klein}, R., \&
  {Sedlmayr}, E. 2001, \aap, 376, 194

\bibitem[{{H{\"o}fner} {et~al.}(2003){H{\"o}fner}, {Gautschy-Loidl}, {Aringer},
  \& {J{\o}rgensen}}]{Hoefner2003A&A...399..589H}
{H{\"o}fner}, S., {Gautschy-Loidl}, R., {Aringer}, B., \& {J{\o}rgensen}, U.~G.
  2003, \aap, 399, 589

\bibitem[{{Hurlburt} {et~al.}(1986){Hurlburt}, {Toomre}, \&
  {Massaguer}}]{Hurlburt1986ApJ...311..563H}
{Hurlburt}, N.~E., {Toomre}, J., \& {Massaguer}, J.~M. 1986, \apj, 311, 563

\bibitem[{{King} {et~al.}(2009){King}, {McCaughrean}, {Homeier}, {Allard},
  {Scholz}, \& {Lodieu}}]{King2009arXiv0911.3143K}
{King}, R.~R., {McCaughrean}, M.~J., {Homeier}, D., {et~al.} 2009, \aap,
  accepted

\bibitem[{{Leggett} {et~al.}(2001){Leggett}, {Allard}, {Geballe}, {Hauschildt},
  \& {Schweitzer}}]{Leggett2001ApJ...548..908L}
{Leggett}, S.~K., {Allard}, F., {Geballe}, T.~R., {Hauschildt}, P.~H., \&
  {Schweitzer}, A. 2001, \apj, 548, 908

\bibitem[{{Leggett} {et~al.}(1998){Leggett}, {Allard}, \&
  {Hauschildt}}]{Leggett1998ApJ...509..836L}
{Leggett}, S.~K., {Allard}, F., \& {Hauschildt}, P.~H. 1998, \apj, 509, 836

\bibitem[{{Ludwig} {et~al.}(2002){Ludwig}, {Allard}, \&
  {Hauschildt}}]{Ludwig2002A&A...395...99L}
{Ludwig}, H.-G., {Allard}, F., \& {Hauschildt}, P.~H. 2002, \aap, 395, 99

\bibitem[{{Ludwig} {et~al.}(2006){Ludwig}, {Allard}, \&
  {Hauschildt}}]{Ludwig2006A&A...459..599L}
{Ludwig}, H.-G., {Allard}, F., \& {Hauschildt}, P.~H. 2006, \aap, 459, 599

\bibitem[{{Ludwig} \& {Nordlund}(2000)}]{LudwigNordlund2000}
{Ludwig}, H.-G. \& {Nordlund}, {\AA}. 2000, in Stellar Astrophysics, ed. K.~S.
  {Cheng}, H.~F. {Chau}, K.~L. {Chan}, \& K.~C. {Leung}, 37

\bibitem[{{Marley} {et~al.}(2007){Marley}, {Fortney}, {Seager}, \&
  {Barman}}]{Marley2007prpl.conf..733M}
{Marley}, M.~S., {Fortney}, J., {Seager}, S., \& {Barman}, T. 2007, in
  Protostars and Planets V, ed. B.~{Reipurth}, D.~{Jewitt}, \& K.~{Keil},
  733--747

\bibitem[{{Nordlund}(1982)}]{Nordlund1982A&A...107....1N}
{Nordlund}, {\AA}. 1982, \aap, 107, 1

\bibitem[{{Robinson} {et~al.}(2003){Robinson}, {Demarque}, {Li}, {Sofia},
  {Kim}, {Chan}, \& {Guenther}}]{Robinson2003MNRAS.340..923R}
{Robinson}, F.~J., {Demarque}, P., {Li}, L.~H., {et~al.} 2003, \mnras, 340, 923

\bibitem[{{Rossow}(1978)}]{Rossow1978Icar...36....1R}
{Rossow}, W.~B. 1978, Icarus, 36, 1

\bibitem[{{Ruiz} {et~al.}(1997){Ruiz}, {Leggett}, \&
  {Allard}}]{Ruiz1997ApJ...491L.107R}
{Ruiz}, M.~T., {Leggett}, S.~K., \& {Allard}, F. 1997, \apjl, 491, L107+

\bibitem[{{Saumon} {et~al.}(2006){Saumon}, {Marley}, {Cushing}, {Leggett},
  {Roellig}, {Lodders}, \& {Freedman}}]{Saumon2006ApJ...647..552S}
{Saumon}, D., {Marley}, M.~S., {Cushing}, M.~C., {et~al.} 2006, \apj, 647, 552

\bibitem[{{Skartlien} {et~al.}(2000){Skartlien}, {Stein}, \&
  {Nordlund}}]{Skartlien2000ApJ...541..468S}
{Skartlien}, R., {Stein}, R.~F., \& {Nordlund}, {\AA}. 2000, \apj, 541, 468

\bibitem[{{Steffen} {et~al.}(1989){Steffen}, {Ludwig}, \&
  {Kr{\"u}{\ss}}}]{Steffen1989A&A...213..371S}
{Steffen}, M., {Ludwig}, H.-G., \& {Kr{\"u}{\ss}}, A. 1989, \aap, 213, 371

\bibitem[{{Stein} \& {Nordlund}(2000)}]{Stein2000SoPh..192...91S}
{Stein}, R.~F. \& {Nordlund}, {\AA}. 2000, \solphys, 192, 91

\bibitem[{{Stein} \& {Nordlund}(2001)}]{Stein2001ApJ...546..585S}
{Stein}, R.~F. \& {Nordlund}, {\AA}. 2001, \apj, 546, 585

\bibitem[{{Stephens} {et~al.}(2009){Stephens}, {Leggett}, {Cushing}, {Marley},
  {Saumon}, {Geballe}, {Golimowski}, {Fan}, \&
  {Noll}}]{Stephens2009ApJ...702..154S}
{Stephens}, D.~C., {Leggett}, S.~K., {Cushing}, M.~C., {et~al.} 2009, \apj,
  702, 154

\bibitem[{{Tsuji}(2002)}]{Tsuji2002ApJ...575..264T}
{Tsuji}, T. 2002, \apj, 575, 264

\bibitem[{{Tsuji} {et~al.}(1996){Tsuji}, {Ohnaka}, \&
  {Aoki}}]{Tsuji1996A&A...305L...1T}
{Tsuji}, T., {Ohnaka}, K., \& {Aoki}, W. 1996, \aap, 305, L1+

\bibitem[{{V{\"o}gler}(2004)}]{Voegler2004A&A...421..755V}
{V{\"o}gler}, A. 2004, \aap, 421, 755

\bibitem[{{Wedemeyer} {et~al.}(2004){Wedemeyer}, {Freytag}, {Steffen},
  {Ludwig}, \& {Holweger}}]{Wedemeyer2004A&A...414.1121W}
{Wedemeyer}, S., {Freytag}, B., {Steffen}, M., {Ludwig}, H.-G., \& {Holweger},
  H. 2004, \aap, 414, 1121

\bibitem[{{Witte} {et~al.}(2009){Witte}, {Helling}, \&
  {Hauschildt}}]{Witte2009A&A...506.1367W}
{Witte}, S., {Helling}, C., \& {Hauschildt}, P.~H. 2009, \aap, 506, 1367

\bibitem[{{Woitke} \& {Helling}(2003)}]{Woitke2003A&A...399..297W}
{Woitke}, P. \& {Helling}, C. 2003, \aap, 399, 297

\bibitem[{{Young}(1998)}]{Young1998JGR...10322775Y}
{Young}, R.~E. 1998, \jgr, 103, 22775

\end{thebibliography}

\end{document}